\documentclass[12pt]{article}

\usepackage{amsmath,amssymb,amsfonts}%
\usepackage{amsthm}%
\usepackage{newtxtext,newtxmath}

\usepackage{graphicx}%
\usepackage{subcaption}  % for subfigures
\usepackage{multirow}%
\usepackage{mathrsfs}%
\usepackage[title]{appendix}%
\usepackage{xcolor}%
\usepackage{textcomp}%
\usepackage{manyfoot}%
\usepackage{booktabs}%
\usepackage{algorithm}%
\usepackage{algorithmicx}%
\usepackage{algpseudocode}%
\usepackage{listings}%
\usepackage{mathtools}
\usepackage{soul}
\usepackage{hhline}
\usepackage{bm}
\usepackage{braket}
\usepackage{makecell} % add this in your preamble
\usepackage{array}      % for m{} column type
\usepackage{tabularx}   % for tabularx (auto column widths)
\usepackage{array,makecell}

\newcommand{\vfix}{\rule{0pt}{2.2ex}\rule[-0.8ex]{0pt}{0pt}}

\newcolumntype{M}[1]{>{\centering\arraybackslash}m{#1}<{\vfix}}

\usepackage{xfrac}
\usepackage[font=footnotesize]{caption}
\usepackage{hyperref}

\newcommand{\Tr}[1]{\mathrm{Tr}\left[#1\right]}

\usepackage[letterpaper,margin=1in]{geometry}

\renewenvironment{abstract}
	{\quotation}
	{\endquotation}

\date{}

\makeatletter
\renewcommand{\fnum@figure}{\textbf{Figure \thefigure}}
\renewcommand{\fnum@table}{\textbf{Table \thetable}}
\makeatother

\usepackage{scicite}
\usepackage{url}

\def\scititle{
	High-dimensional quantum communication with scalable photonic entanglement in time and frequency
}
\title{\bfseries \boldmath \scititle}

\author{
	Kai-Chi Chang$^{1\ast\dagger}$,
	Murat Can Sarihan$^{1\dagger}$,
	Nicky Kai Hong Li$^{2,3\ast\dagger}$,\\
    Florian Kanitschar$^{2,4\dagger}$, Kemal Enes Akyuz$^{1\dagger}$,
    Yujie Chen$^{1}$,
    Dong-Il Lee$^{1}$,\\
    Jin Ho Kang$^{1}$,
    Alwaleed Aldhafeeri$^{1}$,
    Andrew Mueller$^{5,6}$,
    Matthew D. Shaw$^{5}$,\\
    Boris Korzh$^{5}$,
    Maria Spiropulu$^{7}$,
    Paul Erker$^{2,3\ast}$,
    Marcus Huber$^{2,3\ast}$,
    Chee Wei Wong$^{1\ast}$
    \and
	\small$^{1}$Fang Lu Mesoscopic Optics and Quantum Electronics Laboratory, \\[-10pt]
    \small Department of Electrical and Computer Engineering, University of California, Los Angeles, 90095, CA, USA.\and
	\small$^{2}$Atominstitut, Technische Universität Wien, Stadionallee 2, 1020 Wien, Austria.\and
    \small$^{3}$Institute for Quantum Optics and Quantum Information, \\[-10pt]
    \small Austrian Academy of Sciences, Boltzmanngasse 3, 1090 Wien, Austria.\and
    \small$^{4}$AIT - Austrian Institute of Technology, Center for Digital Safety and Security, Giefinggasse 4, 1210 Wien, Austria.\and
    \small$^{5}$Jet Propulsion Laboratory, California Institute of Technology, 4800 Oak Grove Dr., Pasadena, 91109, CA, USA.\and
    \small$^{6}$Applied Physics, California Institute of Technology, 1200 E California Blvd, Pasadena, 91125, CA, USA.\and
    \small$^{7}$Division of Physics, Mathematics and Astronomy, California Institute of Technology, \\[-10pt] 
    \small 1200 E California Blvd, Pasadena, 91125, CA, USA.\and
	\small$^\ast$Corresponding author. Email: uclakcchang@ucla.edu; kai.li@tuwien.ac.at;\\[-10pt]
    \small paul.erker@tuwien.ac.at; marcus.huber@tuwien.ac.at; cheewei.wong@ucla.edu\and
	\small$^\dagger$These authors contributed equally to this work.
}

\begin{document} 

% Insert the title and author list
\maketitle

\newpage

\begin{abstract} \bfseries \boldmath
High-dimensional photonic entanglement holds significant promise for advancing quantum communication, computation, and metrology. For example, large-alphabet quantum communication protocols are known to benefit from enhanced noise resilience and information capacity via multi-bit time-bin encoding. Yet, characterizing high-dimensional entangled states is challenging, as full state tomography becomes prohibitively costly and often requires unrealizable measurements. Here, we demonstrate a scan-free method to characterize high-dimensional entanglement in the time-frequency domain. Our reconstruction achieves a record $5.70\pm0.07$ ebits and a fidelity of $65.4\pm0.4\%$ with the maximally entangled state of local dimension $1021$, certifying the presence of $668$-dimensional entanglement. We further prove the attainability of a secure key rate of $15.6$ kB/s in a composable finite-size,  entanglement-based protocol, and show that in continuous operation, the setup can quickly approach asymptotic key rates. Using commercial telecom components and state-of-the-art low-jitter single-photon detectors, our scalable architecture offers a practical path towards high-rate, noise-resilient quantum communication testbeds.
\end{abstract}
\vspace{10pt}

\noindent

Quantum photonic qudits are a crucial resource for high-dimensional quantum information processing \cite{Flamini_2018,Slussarenko_2019,Friis_2019,Erhard_2020}, environment-resilient quantum key distribution \cite{Pirandola_2020,Xu_2020}, superdense coding \cite{Mattle_1996,Harrow_2004,Barreiro_2008,Hu_2018}, quantum computation \cite{Wang_2018,Alexeev_2021,Arrazola_2021,Chi_2022,Chang_2025}, and quantum imaging \cite{Magana_2019,Moreau_2019}. The availability of large Hilbert space dimensionalities within the photonic degrees of freedom (DoF) – such as frequency-bins \cite{Kues_2017,Joshi_2020,Lu_2022,Clementi_2023,Lu_2023}, time-bins \cite{Tittel_1998,Marcikic_2004,Xie_2015,Martin_2017,Jaramillo_2017,Imany_2019,Chang_2021,Chang_2025_2,Chang_2023,Chang_2024}, temporal modes \cite{Brecht_2015,Fabre_2020,Serino_2023}, orbital angular momentum \cite{Mair_2001,Krenn_2014,Malik_2016,Bavaresco_2018,He_2022,Li_2023,Zia_2023,Scarfe_2025}, path \cite{Wang_2018,Hu_2020,Hu_2021}, and pixel bases \cite{Erker_2017,Schneeloch_2019,Valencia_2020} – enables the encoding of vast amounts of information with fewer photons compared to qubit-based protocols that rely solely on the polarization DoF. However, certifying experimentally generated high-dimensional entangled states is a crucial and challenging task for entanglement in any DoF \cite{Friis_2019,Erhard_2020}. Specifically, the high-dimensionality of these states, such as those produced by the generation of photon pairs, presents an intriguing challenge regarding their measurement \cite{Friis_2019,Erhard_2020}. The number of projective measurements required for full-state tomography (FST) scales quadratically with the dimensionality of the Hilbert space being examined. To tackle this issue, several quantum tomographic methods have been introduced and experimentally demonstrated, such as adaptive tomographic approaches \cite{Mahler_2013,Rambach_2021}, compressed learning \cite{Gross_2010}, mutually unbiased bases (MUB) in the spatial domain \cite{Bavaresco_2018,Erker_2017,Valencia_2020,Brougham_2013}, machine learning \cite{Torlai_2018} and interferometric methods \cite{Martin_2017,Chang_2021,Sahoo_2020}. However, these techniques are either constrained by a priori hypotheses on the quantum state under study \cite{Martin_2017,Chang_2021,Mahler_2013,Rambach_2021,Gross_2010} or by the limited speed and efficiency of the data acquisition \cite{Bavaresco_2018,Erker_2017,Valencia_2020,Torlai_2018,Sahoo_2020}, in the certification of the high-dimensional quantum states. For large-alphabet quantum key distribution (QKD) for example, although proof-of-principle entanglement-based qudit QKD has been examined \cite{Lee_2014,Mirhosseini_2015,Zhong_2015,Liu_2024,Chang_2024_2}, the security relies on many assumptions and is thus not comparable with contemporary qubit implementations, while showing promising signs of potentially high key rates \cite{Zhong_2015,Chang_2024_2}.
Here we address the secure key rate challenge in the specific context of temporally and spectrally correlated biphoton states. We focus on the particular challenge of reconstructing relevant features of the two-photon coincidence postselected quantum state emerging from spontaneous parametric down-conversion (SPDC), specifically in the temporal basis. These quantum states exhibit strong correlations in time and frequency \cite{Erhard_2020}, observed within the plane of biphoton generation, a characteristic also seen in other photon-pair sources based on spontaneous four-wave mixing \cite{Flamini_2018,Slussarenko_2019}.\\

The prevalent method in literature for reconstructing the quantum state emitted by a nonlinear medium relies on local projective techniques \cite{Xie_2015,Jaramillo_2017,Imany_2019,Chang_2021,Chang_2025_2,Chang_2023,Bavaresco_2018,Erker_2017,Schneeloch_2019,Valencia_2020}; this approach suffers from drawbacks related to the measurement times, as it requires successive measurements on non-orthogonal bases and especially in the spatial domain every outcome is associated with either a different filter setting or another detector, rendering the scaling to high-dimensions prohibitively slow or expensive. Here, we introduce a scan-free approach that addresses both issues, offering complete reconstruction of the joint temporal intensity (JTI) of the biphoton state. This information can be visualized by discretizing the arrival time of the biphoton state, defined as the marginals of the coincidence distribution obtained by integrating over the coordinates of one of the biphotons. Then, we can reconstruct the intrinsic JTI of SPDC from post-processing the single measurement. The other measurement is a frequency-resolved JTI, from the time-to-frequency converter: here, such a converter is realized in the commercially available $\pm$ 10,000 ps/nm dispersion emulator and compensator modules with optical loss less than 3 dB. We demonstrate a notable capability, where the straightforward dual-basis measurements allow the retrieval of the joint temporal intensity of the biphoton states in arbitrary temporal modes. In our scheme, the measurement time typically takes only a few seconds, depending on the source brightness, losses in telecom fiber components, and the detection efficiency of single-photon detectors. In contrast, previous projective techniques might require several hours of measurement, even with a smaller subset of modes. \\

With our proposed approach, we first certify the high-dimensional entanglement, both in terms of distillable entanglement and entanglement dimensionality. By employing time-bin encoding and fiber-optic telecom components, in conjunction with our low-jitter single-photon detectors, our results successfully witness up to $5.70\pm0.07$ ebits and 668-dimensional entanglement, both of which are records, considering prior high-dimensional quantum photonic qudit systems \cite{Martin_2017,Chang_2021,Krenn_2014,Bavaresco_2018,Erker_2017,Schneeloch_2019,Valencia_2020}. Our technique also presents dramatically faster measurements (three orders-of-magnitude faster, for 61$\times$61-dimensions, and six orders-of-magnitude faster, for 1021-dimensions, compared to prior works using fidelity bound method \cite{Bavaresco_2018,Erker_2017,Schneeloch_2019,Valencia_2020}), with reliable characterization of biphoton quantum states. For the key throughput challenge, we develop a new scalable semidefinite programming (SDP) based method, capable of certifying composable security against coherent and collective attacks from finite sample sizes. This work thus represents a key step toward realizing a fully scalable high-dimensional quantum photonic platform using the energy-time degree of freedom for a high-rate quantum communication fiber link.\\

%\section*{Results}
\subsection*{High-dimensional photonic qudits in time-frequency from SPDC}

\begin{figure}
\centering
\includegraphics[trim=10pt 5pt 5pt 20pt,width=0.95\textwidth]{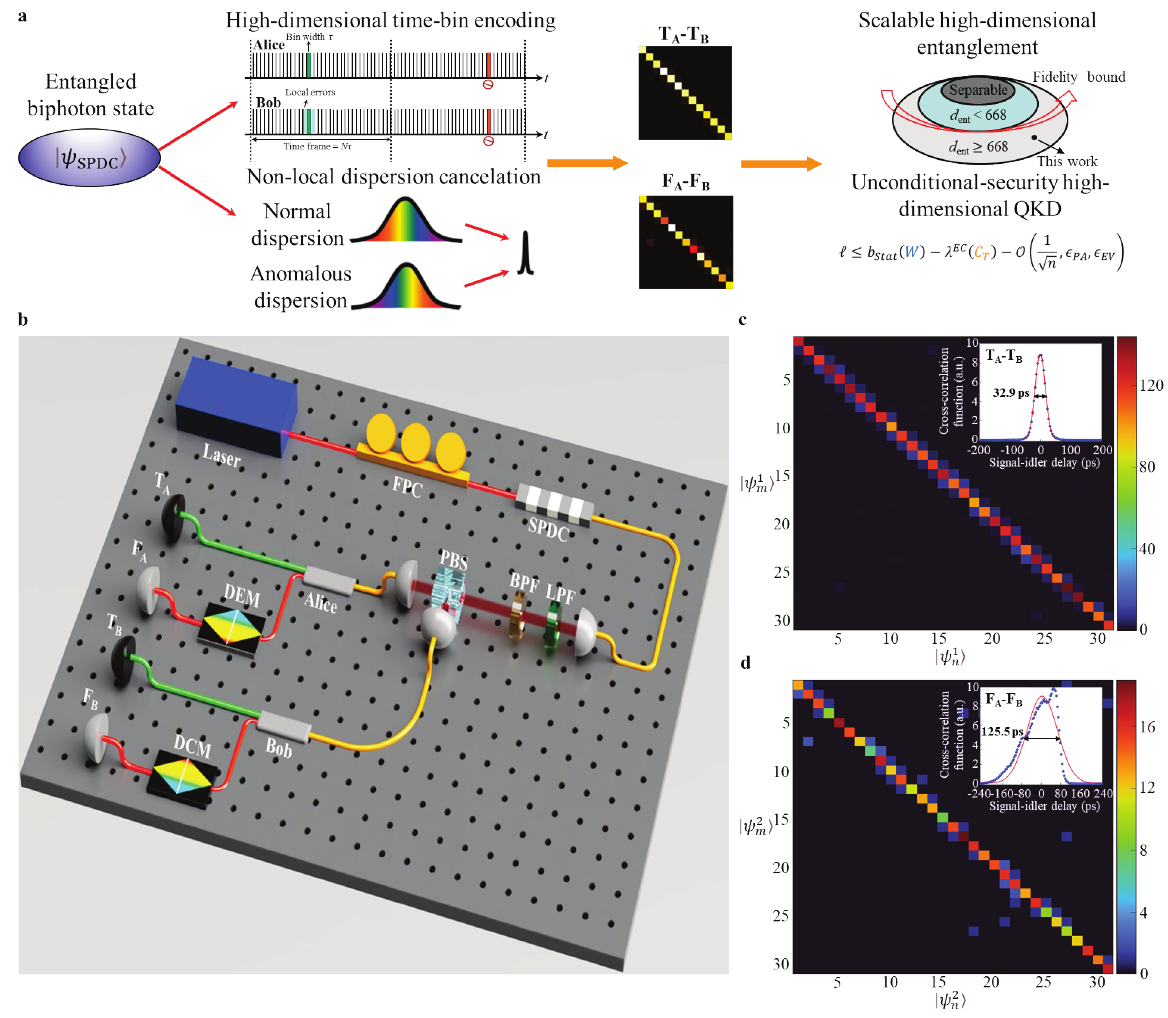}
% \end{center}
\caption{ 
Two-shots measurements for high-dimensional qudit entanglement, high-rate QKD, and 31-dimensional time-frequency resolved joint temporal intensities (JTIs). \textbf{a}, In spontaneous parametric down-conversion (SPDC) photons, the detection of one photon 
fixes the arrival time of the other photon, yielding strong temporal correlations in the JTI. 
We denote temporal measurement basis of Alice and Bob as $T_A-T_B$. By using time-to-frequency convertor, we can perform the frequency-resolved measurements 
in basis $F_A-F_B$. For $T_A-T_B$ the local timing jitter errors are marked as light blue slots, 
while bin-width $\tau$ and number of bins $N$ 
define the time frame length $N\tau$, optimized for the JTI. 
Orange slots indicate there are no registered coincidence photons. For $F_A-F_B$, we utilize non-local dispersion cancelation to recover the narrow temporal correlation and to convert temporal information of SPDC into frequency-domain. \textbf{b}, The experimental setup involves separating signal and idler photons, with Alice and Bob each using 
50:50 fiber beam splitters 
and superconducting nanowire single-photon detectors (SNSPDs) for both $T_A-T_B$ and $F_A-F_B$ measurements.  
\textbf{c}, and \textbf{d}, 
Exemplary 31-dimensional JTIs in $T_A-T_B$ and $F_A-F_B$ bases. The full width at half maximum (FWHM) of temporal correlation peak are 
32.9 ps and  125.5 ps, respectively. For \textbf{d}, we optimize the $F_A-F_B$ measurements by adjusting the pump wavelength. The slight asymmetry of temporal correlation peak comes from the limitation of time-to-frequency convertor. 
Parameters ($\tau$, $N$) are chosen to optimize the JTI: $\tau=200$ ps, $N=1024$ for $T_A$–$T_B$; $\tau=800$ ps, $N=1024$ for $F_A$–$F_B$. The duration of coincidence counting for the data in \textbf{c} and \textbf{d} is 3 seconds, and no accidental subtraction is used.
}
\label{fig:Figure1}
\end{figure}

The high-dimensional Hilbert space is a discretisation of the intrinsic continuous time and frequency correlation of SPDC, where a three-wave mixing process generates the signal and idler photons \cite{Flamini_2018,Slussarenko_2019,Erhard_2020}. Such strong quantum correlations are typically characterized by the JTI and joint spectral intensity measurements \cite{Erhard_2020,Chang_2021,Chang_2025_2,Serino_2023}. Figure \ref{fig:Figure1}a describes the two-shot measurements, enabled by the arrival-time encoding and the time-to-frequency-converter. For the first measurement, we assigned the temporal measurement basis to be $T_A-T_B$, where $T_A$ and $T_B$ are the measurement bases corresponding to the arrival times at Alice and Bob respectively. For the second measurement, we use non-local dispersion cancelation technique to retrieve the narrow temporal correlation of biphoton state and map the temporal measurements into frequency-resolved measurements \cite{Chang_2024_2}, termed $F_A-F_B$. In both measurements, the JTI of SPDC photons can be measured by discretizing the arrival-times of the biphoton state. This process involves high-dimensional temporal encoding, and we define the marginals of the coincidence distribution by integrating over the coordinates of one of the biphotons. Local timing jitter errors are light blue slots, and there are two key parameters to optimize the JTI: bin-width $\tau$ and the number of bins $N$. The time frame length is hence defined as product of bin-width $\tau$ and the number of bins, $N$. Orange slots indicate there are no registered coincidence photons. We note that the JTI measurements reconstructed by this study are dimensionally-independent from the large-alphabet encoding nature of arrival-times. The JTI measurements are only limited by the detectable coincidence counts from the experimental quantum photonic platforms. With this approach, we can reconstruct the JTIs of SPDC by post-processing the two-shot measurements: $T_A-T_B$ and $F_A-F_B$ form the two approximate measurement mutually unbiased bases  for evaluating high-dimensional entanglement witnesses and proving security parameters for QKD.

\subsection*{Experimental setup and measured mutually-unbiased bases} 
Building on the principle detailed in the previous section, we experimentally developed a platform utilizing large-alphabet time-bin encoding and time-to-frequency converter to reconstruct the biphoton state. This biphoton quantum state is emitted via SPDC in a type-II process, where the energy-time entangled photon-pairs generated from our continuous-laser-pumped nonlinear $\chi^{(2)}$  waveguide is expressed in the time-domain as: 
\begin{equation}
  \varphi_{\text{biphoton}} \propto \int  dt_- \,\varphi_{\text{biphoton}}(t_-) \vert t_+ + t_-\rangle_{Signal} \otimes \vert t_+ - t_-\rangle_{Idler} ,        
\end{equation}
where $t_+ =(t_{Signal}+t_{Idler})/2$, and $t_- = (t_{Signal} - t_{Idler})/2$. $\varphi_{\text{biphoton}}(t_- )$ is the joint temporal amplitude, and its magnitude square $ \vert \varphi_{\text{biphoton}}(t_- ) \vert^2$ is the JTI of $t_-$. This JTI of biphoton is known to be difficult to measure in energy-time DoF, often due to the limitation of detection jitter \cite{Flamini_2018,Slussarenko_2019,Erhard_2020}. \\

A visual representation of the experimental setup for the high-dimensional arrival-time encoding to discretize the JTI is shown in Figure \ref{fig:Figure1}b (see Methods for more details \cite{Korzh_2020,Colangelo_2023}). With the SPDC-generated photon pairs, the continuous-wave filtered, with the entangled signal and idler photons separated by a polarization beam splitter. Both Alice and Bob utilize their 50:50 fiber beamsplitters for conducting biphoton temporal correlation measurements ($T_A-T_B$) and frequency-resolved correlation measurements ($F_A-F_B$). Each side employs two low-jitter 
SNSPDs for detection. Figure \ref{fig:Figure1}c inset depicts the measured cross-correlation of biphotons in a temporal basis ($T_A-T_B$) using two SNSPDs with low-jitter. In this temporal measurement basis, the second-order correlation peak has a full-width half-maximum (FWHM) measured at $\approx$ 32.9 ps, constrained by the detector and electronic jitter within our coincidence counting module. For the frequency-resolved measurements basis $F_A$ and $F_B$, we insert a pair of time-to-frequency converters of $\pm$ 10,000 ps/nm dispersion emulator and compensator modules (DEM and DCM), with the optical loss less than 3 dB. Via non-local dispersion cancellation \cite{Liu_2024,Chang_2024_2}, we retrieve the narrow correlated temporal peak with FWHM of about 125.5 ps, bounded by the detectors and the dispersion modules we used, as shown in inset of Figure \ref{fig:Figure1}d. The effective frequency-resolution in this measurement is obtained as FWHM timing jitter normalized by the applied dispersion, corresponding to $\approx$  0.00329 nm (0.41 GHz), sizably smaller than our SPDC source FWHM bandwidth of $\approx$  250 GHz. For our measurements in Figure \ref{fig:Figure1}d, we optimize the frequency-resolved $F_A$ and $F_B$ measurements by adjusting the pump wavelength, and the slight asymmetry of temporal correlation peak comes from the imperfection of our dispersive components. With both the temporal and frequency correlated bases, subsequently we can capture the arrival-time stamps of coincidences originating from these two-shot measurements. Figures \ref{fig:Figure1}c and d show an example of the resulting discretized 31-dimensional large-alphabet JTIs in both $T_A-T_B$ and $F_A-F_B$ . measurement bases. For $T_A-T_B$ basis in Figure \ref{fig:Figure1}c, we choose a 200 ps bin-width $\tau$ and the number of bins $N$ at 1024; for the $F_A$ and $F_B$ basis in Figure \ref{fig:Figure1}d, while we use bin-width $\tau$ of 800 ps and the number of bins $N$ of 1024 for $F_A$ and $F_B$ basis. These parameters are chosen to optimize the JTIs in both the time and frequency basis. Each coincidence counting of the large Hilbert space is completed within 3 seconds and no accidental subtraction is used.\\[-5pt]

We then conducted the validation of mutual unbiasedness of the two bases by employing cross-basis measurements within our time-frequency bases. Two $d$-dimensional bases, denoted by $m$ and $n$, are considered mutually unbiased when their constituent elements, denoted by $i$ and $j$, adhere to the following relation \cite{Bavaresco_2018,Erker_2017}:
\begin{equation}
\vert \langle t_{i}|d_{j}\rangle \vert^2 = \vert \langle \psi_{m,i}|\psi_{n,j} \rangle \vert^2 =
\begin{cases}
      \tfrac{1}{d} \,\,\,\quad \text{for} \quad  m \neq n \\
       \delta_{ij} \quad \text{for} \quad  m = n 
\end{cases}      
\end{equation}                        
for all $i$ and $j$. $t_{i}$ is the temporal basis, and the $d_{j}$ is the dispersive basis that is conjugate to the temporal basis. One should note here that the two bases span overlapping, but not identical Hilbert spaces $\sum|t_i\rangle\langle t_i|\neq\sum|d_i\rangle\langle d_i|$. We should thus preface that all measurements are made under a double-fair sampling assumption: for one, we assume coincidences to be representative of the whole ensemble, even though singles are ignored and the correlations in temporal and frequency bases are, on average, representative of the correlations therein, despite only sampling from smaller subspaces. The fact that the coincidences lead to MUB consistent results still implies that the results obtained in one base give minimal information about the corresponding outcomes in the other basis. We verify their unbiased nature by measuring cross-detection probabilities. This involves utilizing our cross-basis time-frequency and frequency-time measurements. Our verification results are given in Figure \ref{fig:S1} of the Supplementary Materials. The normalized Frobenius norm of the difference between the normalized time-frequency bases' correlation matrix and the ideal correlation matrix for MUBs in $d=1021$ is $0.05\%$. We summarized our cross-basis verification results for various dimensions, in which we certify entanglement
and evaluate secure key rates, in Table \ref{tab:FrobeniusNorms} of the Supplementary Materials. \\

\begin{figure}[h!]
\centering
\includegraphics[trim=32pt 5pt 32pt 0, width=0.7\textwidth]{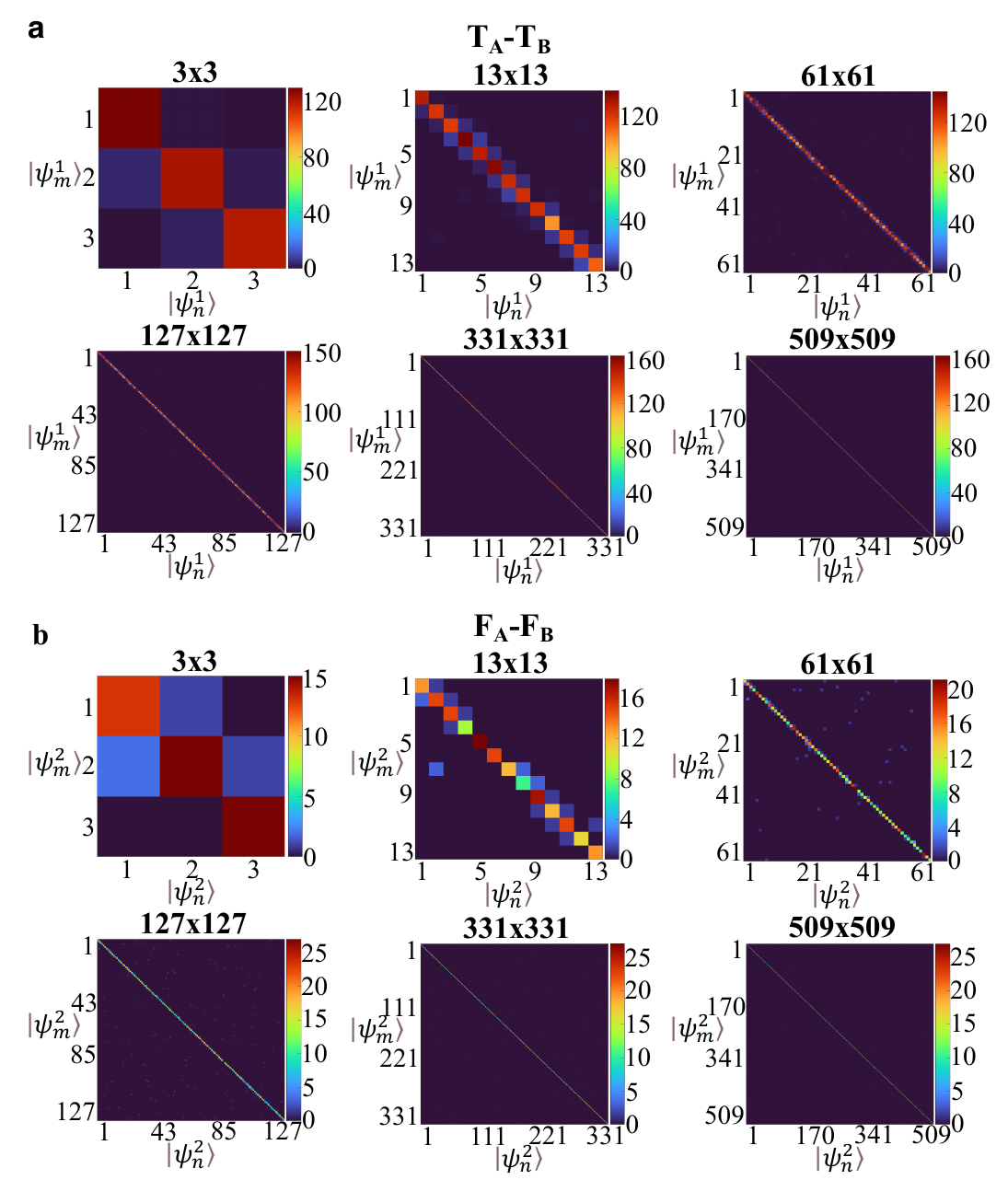}
% \end{center}
\caption{Experimental time-frequency bases up to 509$\times$509 dimensions.  \textbf{a}, and \textbf{b}, An experimental measured 3$\times$3, 13$\times$13,  61$\times$61, 127$\times$127, 331$\times$331,  and 509$\times$509 Hilbert space dimensional JTI for temporal and frequency-resolved measurement basis. We can observe that our JTIs from both bases are indeed dimensionally independent with respect to the measurements, owing to the large-alphabet arrival-time encoding, and the sufficient detected coincidence counts in our experimental setup. For all the measurements presented here, the JTI of $T_A-T_B$ basis has higher diagonal coincidence counts than that of $F_A-F_B$ basis, which is mainly due to the losses in of the time-to-frequency converter (which is in total of $\approx$ 6 dB). From the same reason, we observe that the $F_A-F_B$ basis is noisier than the $T_A-T_B$ basis. For all experimental data in \textbf{a} and \textbf{b}, the duration of measured coincidence counting is 3 seconds, and the raw data is presented here. 
}
\label{fig:Figure2}
\end{figure}

Figure \ref{fig:Figure2} shows the coincidence measurement outcomes in the experimental time-frequency bases up to 509$\times$509-dimensions, measured with bin-width $\tau$ of 800 ps and number of bins $N$ of 1024 in the two-shot measurements, akin to Figures \ref{fig:Figure1}c and d. We illustrate here for experimental Hilbert spaces measured at prime numbers 3$\times$3, 13$\times$13, 61$\times$61, 127$\times$127,  331$\times$331, and 509$\times$509. We observe that our JTIs from both bases are scalable in measurement dimensions, with the dimensional measurement independence due to our large-alphabet arrival-time encoding approach, given that we have sufficient detected coincidence counts in our experimental setup, even up to 509$\times$509 spaces. For all the measurements presented here, the JTI of $T_A-T_B$ basis has higher diagonal coincidence counts than that of $F_A-F_B$ basis, due to the $\approx$ 3 dB losses in each time-to-frequency dispersion module, and thus the $F_A-F_B$ basis is noisier. \\

\subsection*{Certification of high-dimensional entanglement}

In this section, we certify and quantify the high-dimensional entanglement described by our experimental data that come from the scalable and scan-free JTI measurements in our scheme. Our approach is to 
utilize a fidelity-based Schmidt-number witness from \cite{Bavaresco_2018} and a distillable entanglement bound from \cite{BertaChristandlColbeckRenesRenner2010,DevetakWinter2005} (see also Eq.\;(17.135) in~\cite{BertlmannFriis2023}) that requires measurements in (at least) two complementary bases. We apply these methods to the data obtained from our temporal and dispersive basis measurements. Let us first give a brief overview of both approaches, and defer the more technical summary to the Supplementary Materials.\\

The first method \cite{Bavaresco_2018} certifies the Schmidt number of a state $\rho$ by estimating a fidelity lower bound $\tilde{F}(\rho,\Phi)$ with respect to a pure target entangled state $\ket{\Phi}$ (which we choose to be the maximally entangled state) with the maximum Schmidt rank $d$. If $\tilde{F}(\rho,\Phi)$ exceeds the upper bound $\mathcal{B}_k$, which we define in the Supplementary Materials, for all states with Schmidt number $k$, then $\rho$ is certified to have Schmidt number at least $k+1$. In the following, the Schmidt number or the \textit{entanglement dimension} $d_{ent}$ of a state $\rho$ will only be referred to the maximum Schmidt number that we can certify from $\rho$. Intuitively, a higher entanglement dimension $d_{ent}$ enables more information to be encoded and transmitted securely (as we will see in the next section), making it a natural quantifier of high-dimensional entanglement.\\

The second method \cite{BertaChristandlColbeckRenesRenner2010,DevetakWinter2005} lower bounds the \textit{distillable entanglement} or \textit{entanglement of distillation} ($E_D$), which represents the maximum asymptotic average number of maximally entangled two-qubit states that can be extracted per copy of a quantum state $\rho$ using classical communication and local operations \cite{BennettBrassardPopescuSchumacherSmolinWootters1996,Martin_2017}. For a pair of two-dimensional quantum systems, the maximum entanglement they can have is 1 ebit, which corresponds to a Bell state. In contrast, higher-dimensional systems can potentially contain up to $\text{log}(d)$ bits of entanglement, thereby enabling certification in high-dimensional scenarios.
The bound uses the respective conditional Shannon entropies of the measurement outcomes in the first and second bases, as well as the maximum overlap between the two bases (which would be $1/d$ in the case of ideal MUBs, as presented in the prior section). With limited counts, one expects individual elements to deviate statistically from the mean, thus rendering the determination of the maximum overlap of the two bases a challenge. We work with the hypothesis of mutual unbiasedness, which we test and see that the expected deviation is in line with purely statistical fluctuations around the mean (more details are provided in the Supplementary Materials).\\

\begin{figure}
\centering
\includegraphics[width=0.925\textwidth]{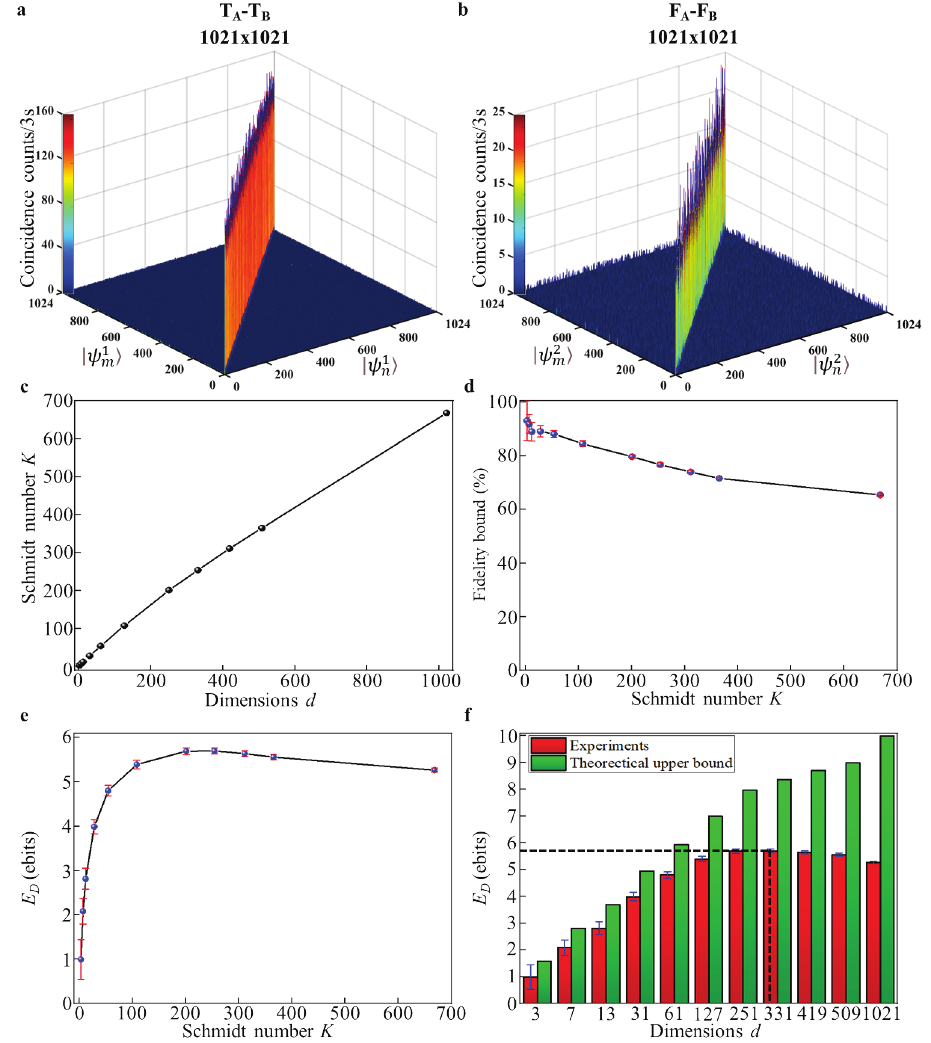}
% \end{center}
\caption{High-dimensional entanglement certification from a maximum of 1021-dimensional JTI measured in the temporal and frequency bases. \textbf{a} and \textbf{b}, 
Coincidence counts for measuring in the $T_A\!-\!T_B$ and $F_A\!-\!F_B$ bases in a two-shot setting, with consistent bin width $\tau$ and number of bins $N$ as shown in Figures \ref{fig:Figure1} and \ref{fig:Figure2}. The 
timing for coincidence counting is 3 seconds, and the 
data are reported without accidental subtractions. 
The strong 
correlations in both measurement bases 
signify robustness of large-alphabet arrival-time encoding. 
\textbf{c} and \textbf{d}, 
The certified Schmidt number $k$ for each measured local dimension $d$, and their associated lower bound of the fidelity $\tilde{F}(\rho,\Phi)$ with respect to the maximally entangled state are shown. The maximum certified Schmidt number is 668 at a fidelity of $65.4\pm0.4\%$ in $d=1021$. 
\textbf{e} and \textbf{f}, The distillable entanglement $E_D$ for various local dimensions $d$ are shown together with the certified Schmidt numbers and the theoretical upper bound of $E_D$, $\text{log}_2(d)$.}
\label{fig:Figure3}
\end{figure}

Before we present the entanglement certification results, let us first demonstrate the scalability of our JTI measurements performed in the time-frequency bases. In Figures \ref{fig:Figure3}a and b we present the biphoton coincidence counts from 1,021$\times$1,021-dimensional discretized JTI measurements of the time- and frequency-resolved measurement basis, with consistent bin-width $\tau$ and number of bins $N$ as described in Figures \ref{fig:Figure1} and \ref{fig:Figure2}. Even up to a local Hilbert space dimension of 1021, strong quantum temporal correlations are observed in both measurement bases. This represents the robustness of large-alphabet arrival-time encoding, and such measurements are obtained in a two-shot setting. Note that in both Figures \ref{fig:Figure3}a and b, the duration of measured coincidence counting is 3 seconds, and the raw experimental data is presented without any accidental subtractions. \\

We now move on to discuss our entanglement certification results for different input dimensions, which are shown in Figures~\ref{fig:Figure3}c{\textendash}f.
From the full 1,021$\times$1,021-dimensional Hilbert space corresponding to the discretized JTI measurements, 
we can certify an entanglement dimension $d_{ent}$ up to 668 and a distillable entanglement of $E_D=5.27\pm 0.04$ ebits with the two-shot time-frequency bases measurements. 
Interestingly, the maximum distillable entanglement $E_D$ of $5.70\pm0.07$ ebits is achieved when the entangled dimension $d_{ent}$ is 246 (where the local dimension $d$ is 331). We attribute the discrepancy in the local dimensions at which the two quantities achieve maximum to their different noise sensitivity, as higher-dimensional measurements tend to be noisier. We also compare our experimental certified $E_D$ with the theoretical upper bound of $\text{log}(d)$ in Figure~\ref{fig:Figure3}f and observe the same falling behaviour in $E_D$, which suggests noise in the two-shot JTI measurements as $d$ grows. 
We further support this observation with Figure~\ref{fig:Figure3}d, where we show that the lower bound on the state fidelity $\tilde{F}(\rho,\Phi)$ reaches the minimum at $65.4\pm0.4\%$ for $d=1021$. 
The uncertainty in the fidelity is calculated based on the assumption that the measured frequency of each measurement outcome is the mean value of a Poisson distribution. By sampling these distributions of all outcomes jointly for 1000--2000 times and assuming a final Gaussian distribution for the computed fidelities, the error bars are taken to be 3 standard deviations from the observed fidelity. 
We remark that both the certified entanglement dimension $d_{ent}$ and distillable entanglement $E_D$ are record measurements to date. For a comprehensive comparison with known results, please refer to Table~\ref{fig:Table1}.\\
\begin{table}[th!]
\centering
\renewcommand{\arraystretch}{1.2}
\resizebox{\textwidth}{!}{%
\begin{tabular}{|
  M{2.3cm}|| M{1cm}| M{0.9cm}| M{2.3cm}| M{1.5cm}|
  M{1cm}| M{1.2cm}| M{0.9cm}| M{0.9cm}|
  M{1.4cm}| M{2cm}| M{1.1cm}| M{1.4cm}|
}
\hline
Experiments & \cite{Krenn_2014} & \cite{Erker_2017} & \cite{Steinlechner_2017} &
\cite{Martin_2017} & \cite{Bavaresco_2018} & \cite{Schneeloch_2019} &
\cite{Hu_2020} & \cite{Valencia_2020} & \cite{Chang_2021} &
\cite{Lu_2022} & \cite{Korzh_2020} & This work\\
\hline\hline
Domains & OAM & Pixel & \makecell{Energy-\\time-\\polarization} &
\makecell{Energy-\\time} & OAM & Pixel & Path & Pixel &
\makecell{Energy-\\time} & Frequency & \makecell{Time-\\bin} &
\makecell{Energy-\\time} \\
\hline
Years & 2014 & 2017 & 2017 & 2017 & 2018 & 2019 & 2020 & 2020 & 2021 & 2022 & 2025 & 2025\\
\hline
\makecell{Certified\\entangled\\dimensions} & 100${}^\text{[1]}$ & 3 & 4 & 18 & 9 & 10 & 32 & 97 & 4 & 8 & 16 & 668\\
\hline
$E_D$ (ebits) & N/A & 3.05 & 1.47 & 4.1${}^\text{[2]}$ & N/A & 3.43${}^\text{[3]}$ &
3.73 & 4.0 & 1.89${}^\text{[2]}$ & 2.32 & 1.992 & 5.701\\
\hline
\end{tabular}}
\caption{Comparison of assumptions used in various high-dimensional quantum photonic experiments. $[1]$: Conservation of OAM. $[2]$: No cross-talk in the computational basis. $[3]$: Raw data, without accidental subtraction. N/A: information not available or not applicable.}
\label{fig:Table1}
\end{table}

The results in Figures \ref{fig:Figure1}{\textendash}\ref{fig:Figure3} clearly demonstrate the advantage of our approach, which is attributable to the following novel techniques. Employing SNSPDs with reduced timing-jitter in the telecom wavelengths \cite{Korzh_2020,Colangelo_2023} allow us to utilize smaller bin-widths $\tau$, such that we are able to encompass the entire temporal correlation peak. Simultaneously, we can steadily increase the number of bins $N$ to expand the dimensionality of our discretized JTIs. We note that the optimal parameters for JTIs presented in this work can be adapted to other quantum photonic systems by considering the corresponding FWHM of temporal correlation peaks, coincidence counts of SPDC source, loss of telecom components, efficiency, and timing-jitter of the detectors. This multi-bit temporal encoding scheme ensures that the number and duration of measurements 
remain constant over the subspace dimension for our discretized JTIs. 
We illustrate this feature across various temporal subspaces in Figures \ref{fig:Figure2} and \ref{fig:Figure3}. Therefore, our approach also offers significantly faster measurements, with an acquisition time three orders of magnitude faster for 61-dimensional data and six orders of magnitude faster for 1,021-dimensional data. We summarize the comparison of the required number of local projective measurements versus different dimensions $d$ for various techniques in Methods. Additionally, having smaller bin-widths $\tau$ and a larger number of bins $N$ is preferable for achieving a higher key capacity in temporal encoding with large alphabets \cite{Zhong_2015,Liu_2024,Chang_2024_2}. \\

\subsection*{Large-alphabet quantum key distribution}

After successfully generating and certifying high-dimensional entanglement within our time-frequency (approximate) MUBs, we demonstrate one of the key applications of quantum photonic qudits: large-alphabet \cite{Lee_2014,Mirhosseini_2015, Zhong_2015, Liu_2024, Chang_2024} quantum key distribution \cite{Bennett_Brassard_1984, Ekert_1991}. Indeed, transmitting delicate quantum correlations through a noisy channel poses a significant hurdle in quantum communication tasks \cite{Pirandola_2020, Xu_2020, Usenko_2025}. High-dimensional QKD protocols address this by encoding dense information in entangled biphoton states, enabling high key throughput \cite{Zhong_2015, Chang_2024} with enhanced robustness against detector dead-time and environmental noise \cite{Chang_2024, Mirhosseini_2015, Zhong_2015, Liu_2024, Kanitschar_2024}.  
Different trust models exist for QKD \cite{Pirandola_2020, Scarani_2009, Usenko_2025}, ranging from fully device-trusted prepare-and-measure schemes to device-independent protocols that rely on loophole-free Bell violations. Entanglement-based QKD represents a reasonable balance between these extremes: it provides security against coherent attacks, is readily certifiable, avoids the need for specialized countermeasures such as decoy-state methods, and still achieves competitive key rates in realistic implementations.\\

By combining recent advances in high-dimensional protocols and coherent composable finite-size security proofs, originally developed for Franson certified time-bin experiments, we generalise the protocol and provide the first comprehensive security analysis of high-dimensional, finite-size protocols that are based on two mutually unbiased bases. Crucially, the actual phase relation between the two bases does not need to be known or assumed, only the relative overlap. This is inherited from the witness used in the security proof, which is invariant under relative phase transformations. For the overlaps, we performed cross-basis measurement, certifying a good correspondence of the ideal positive operator-valued measure (POVM) with the experimental implementation, subject to a fair sampling assumption. In more detail, we use the measured coincidence-click matrices to determine the observed average of the observable $\hat{W} = \sum_{i=0}^{d-1} A_F^i \otimes B_F^i$ with $\{A_F^i\}_i$ and $\{B_F^i\}_i$ being Alice's and Bob's frequency basis measurements, respectively. In line with the assumptions of device-dependent QKD, we assume that the measurement devices are in Alice's and Bob's trusted laboratories, hence under their control. In particular, this means we assumed that the theoretical POVM elements are also practically implemented (up to relative phases). While a good alignment between the theoretical description and the practical implementation could be experimentally verified, quantifying deviations between the theoretical model and the practical implementation is still an active area of research, even for qubit-based protocols \cite{Tupkary_2025, CurrasLorenzo_2025}. Thus, although beyond the scope of the present proof-of-principle work, bounding the influence of small deviations from theoretical measurements remains an interesting avenue for future research. Based on our witness-based approach, as we detail in the Supplementary Materials, we certify a record asymptotic key rate using only short measurement times and statistics. Additionally, we use a variable-length security argument \cite{Kanitschar_2025b} to demonstrate the potential for a composable finite-size secure key rate in realistic measurement times. \\

\begin{figure}
\centering
\includegraphics[trim=10pt 0 5pt 10pt, width=1\textwidth]{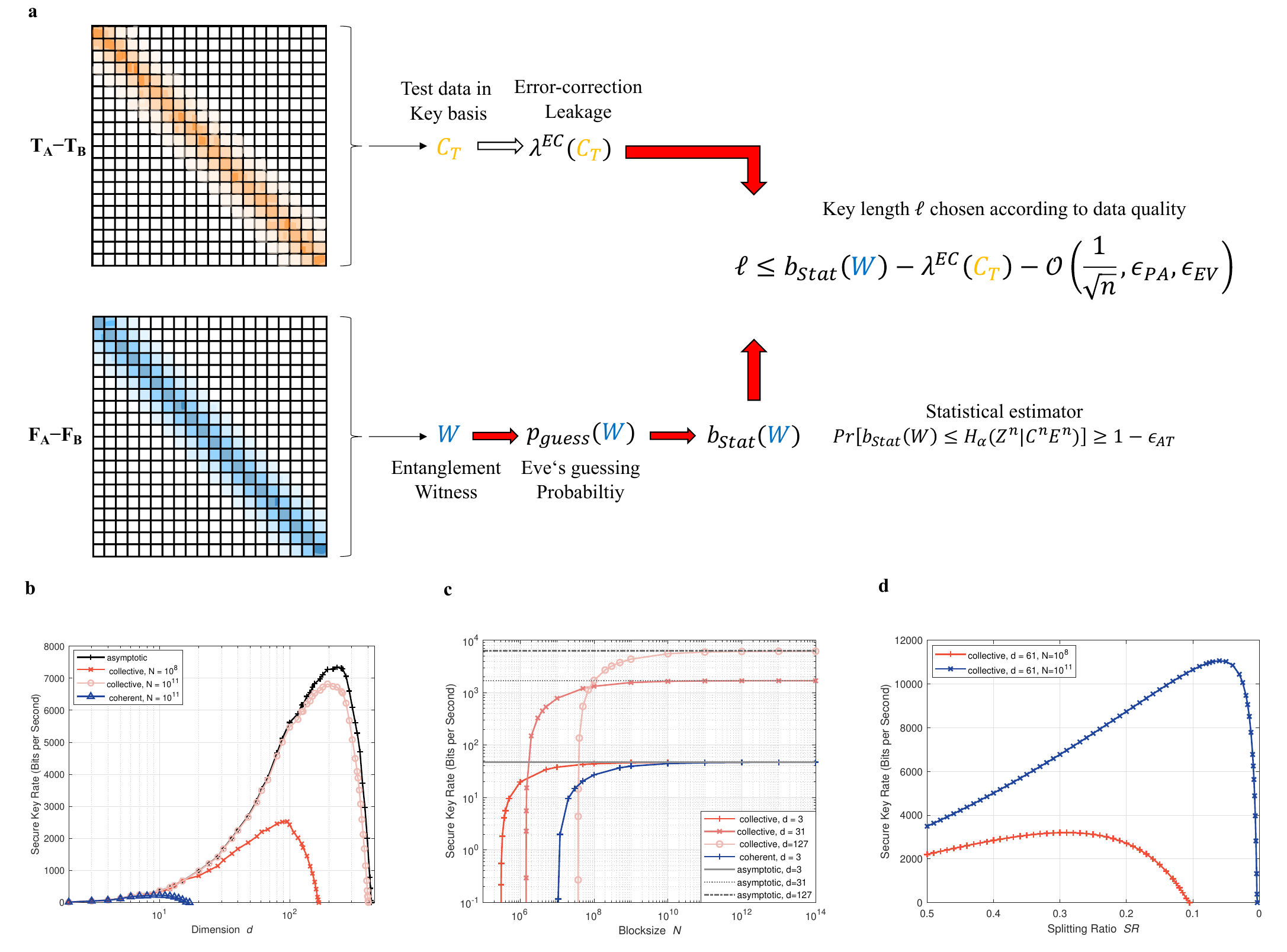}\\
\caption{\textbf{a} Illustration of the security argument: Based on the recorded coincidence-click matrices, we derive two quantities. First, based on the FF-clicks, we derive the expectation of an entanglement witness, which allows us to bound Eve's guessing probability on the final key, which, in turn, allows us to derive a statistical estimator. That is a high probability lower-bound on the private entropy of the key given Eve's side information. Second, from the disclosed TT-clicks, we derive the error-correction leakage. Subtracting the error-correction leakage from the statistical estimator yields (up to second-order correction terms) a reliable high-probability lower bound on the secure key length. \textbf{b} Secure key rate in bits per second versus system dimension in four different scenarios: asymptotic (black pluses), i.i.d. collective attacks with block size $N=10^{8}$ (orange crosses), i.i.d. collective attacks with block size $N=10^{11}$ (peach circles), and coherent attacks with $M=10^{11}$ (blue triangles). Based on our data, we obtain optimal system dimensions of $d_{\mathrm{opt}}^{\mathrm{asym}}(\infty)=232$ for the asymptotic scenario, $d_{\mathrm{opt}}^{\mathrm{coll}}(10^{8})=96$ for i.i.d. collective attacks with $N=10^8$, $d_{\mathrm{opt}}^{\mathrm{coll}}(10^{11})=196$ for i.i.d. collective attacks with $N=10^{11}$, and $d_{\mathrm{opt}}^{\mathrm{coh}}(10^{11})=9$ for coherent attacks with $N=10^{11}$. \textbf{c} Secure key rate in bits per second versus block size $N$ for three different dimensions. In all three cases, the curves converge to the asymptotic rates (horizontal lines) already for practically viable block sizes. \textbf{d} Examination of the secure key rate in bits per second versus splitting ratio between time and frequency measurements for two different block sizes and fixed dimension $d=61$. For $N=10^{8}$, we find an optimal splitting ratio of $29\%$, while for $N=10^{11}$, the optimal splitting ratio is found to be $6\%$. This highlights that optimal splitting is far below the default $50\%$ and the key rates can be improved significantly by optimising this parameter.\label{fig:IllustrationSecurity}}
\end{figure}

While fixed-length security arguments, which are predominant in the literature and build security around the expected behavior of the quantum channel connecting Alice and Bob, we follow an adaptive-length approach and build the security argument around the observed statistics. For fixed-length approaches, the expected channel behavior needs to be fixed before the protocol execution. After the protocol run, one performs an acceptance test, where the observed statistics are compared to the pre-defined expected behavior. In case the test accepts, the protocol produces a key of a fixed length; otherwise, the protocol aborts and does not produce key at all. This is quite restrictive and, in practice, leads to an excessive amount of aborted rounds. We circumvent this problem and build our security argument around the observed statistics. In more detail, we adapt the variable-length approach \cite{Tupkary_2024} for HD-QKD protocols from Ref. \cite{Kanitschar_2025b}. However, we replace the witness-inspired completion technique \cite{Kanitschar_2025} by data obtained from mutually unbiased basis measurements. Thus, we directly observe the correlation between Alice's and Bob's test rounds in the frequency basis $W = \sum_{i=0}^{d-1} P(ii| \textrm{FF})$, where $P(ii|\textrm{FF})$ is the probability that Alice and Bob obtain equal outcomes when both measure in the frequency basis \cite{Doda_2021}. Based on this observation, we find a statistical estimator $b_{\mathrm{stat}}(W)$, which is a high-probability lower bound on the private entropy given Eve's side information. The details of the security argument are illustrated in Figure \ref{fig:IllustrationSecurity}a. For further details, we refer to the Supplementary Materials.\\[-15pt]

We use this novel approach to illustrate the key rate potential of our setup. We note that, in line with existing works on the implementation of quantum key distribution protocols, our analysis conditions on coincidence clicks. Under the fair sampling assumption, those conditioned rounds are representative of the full $N$-round quantum state. While this remains an assumption in the present work, which is a proof of principle, it does not represent a fundamental obstacle and can be removed in future works. Besides conditioning on coincidence clicks, we do not perform any additional postselection (e.g., accidental subtraction), which, in principle, could further enhance the key rate but would require careful treatment in the security analysis.  In Figure \ref{fig:IllustrationSecurity}b, we plot the secure key rate per second over the dimension of the underlying quantum system. For the asymptotic data, we observe a peak for $d=232$, while for $N=10^{11}$ the inferred collective attack key rate peaks only slightly below at $d=196$. Even for $N=10^{8}$, we observe a peak for $d=96$ and therefore a clear outperformance of high-dimensional QKD versus qubits. This holds even true for coherent attacks. As we illustrate in Figure \ref{fig:IllustrationSecurity}c, our finite-size rates approach the asymptotic limit already for relatively small and therefore realistic block sizes. All results so far referred to our standard setting, where Alice and Bob measure in both bases with equal probability. However, as we argue in Figure \ref{fig:IllustrationSecurity}d, this is far from optimal. Optimising this splitting ratio can boost the key rates further, as showcased for blocksize $N=10^{11}$, where the key rates can be increased by a factor of $3$ compared to $50:50$ splitting. Optimising over both dimension and splitting ratio, we certify a composable collective i.i.d. key rate potential of $15.6$ kB/s for $d=232$ with a splitting ratio of $14\%$. The primary goal of this section was to demonstrate the key rate potential of the platform and measurement method in use, and to provide a comparison with qubit protocols. This included optimisations over the chosen dimension and the splitting ratio, as well as the examination of the key rate for two different levels of security and across different block sizes. While those considerations are essential for understanding the behavior and the key rate potential of the setup, for continuous operation, one would usually fix those parameters before protocol execution based on the hardware characteristics (e.g., source brightness, post-processing capabilities) and an estimate of the channel behavior (loss and noise). However, since the expected characteristics of the whole system do not enter the adaptive security argument, deviations thereof do not compromise security.

\subsection*{Conclusion}

In this study, we presented a novel approach for reconstructing the temporal structure of correlated biphoton quantum states. Our proposal leverages the large-alphabet time-bin encoding of SPDC photon-pairs and utilizes low-jitter 
single-photon detectors to probe the arrival time of the qudit states. Only two measurements are required to reconstruct the JTI of the biphoton entangled states with high fidelity. We concentrated on the case of SPDC, generated from a nonlinear waveguide, analyzed the temporal and frequency-resolved correlations, high-dimensional energy-time entangled biphoton states, and large-alphabet QKD in a telecom fiber link. The results demonstrate the superiority of this technique over projective techniques (such as in \cite{Bavaresco_2018,Erker_2017,Schneeloch_2019,Valencia_2020}) for benchmarking highly correlated quantum states. We observe that performing a projective measurement on an 1,021-dimensional subspace, would require several days to accumulate the necessary statistics for $1,021^2$ (or $\approx2^{19.99}$) projections. This extended duration is due to the low count rates associated with the lossy techniques used for mode projection. In contrast, our approach enables us to gather the required data within a few seconds, regardless of the subspace dimensionality being analyzed (with the only limitation being the detectable coincidence counts in our experimental system).  
We certified 668-dimensional entanglement at $d=1021$ and distillable entanglement of up to $5.70\pm0.07$ ebits at $d=331$, through maintaining high fidelities with the maximally entangled state states with a minimum of $65.4\pm0.4\%$ for $d=1021$. \\

Here, in addition to high-dimensional entanglement certifications, we extended recent QKD protocols to demonstrate an exemplary quantum communication experiment. We developed a composable finite-size security proof tailored towards the two measurements and based upon \cite{Doda_2021}, proving the capacity for a secret key of $15.6$ kB/s. Our adaptable approach employs optical fiber components commonly used in telecom wavelengths, alongside the recent low-jitter \cite{Colangelo_2023} 
single-photon detectors. These numbers can be further improved by the continuous advancement of SPDC sources, telecom fiber components, low-jitter \cite{Korzh_2020,Colangelo_2023}, and highly-efficient SNSPDs. \\

Besides the high-dimensional time-bin encoding, another key ingredient is to generate frequency-resolved temporal correlations with time-to-frequency converters. Future studies will focus on extending this platform to various biphoton and multiphoton states, produced from separate distance sources. The results could pave the way for scalable high-dimensional quantum information processing as well as robust high-rate quantum communication networks, towards the fully deployable quantum internet.

\bibliography{science_template} % for a file named science_template.bib

\begin{thebibliography}{10}
\providecommand{\url}[1]{\texttt{#1}}
\expandafter\ifx\csname urlstyle\endcsname\relax
  \providecommand{\doi}[1]{doi:\discretionary{}{}{}#1}\else
  \providecommand{\doi}{doi:\discretionary{}{}{}\begingroup \urlstyle{rm}\Url}\fi

\bibitem{Flamini_2018}
F.~Flamini, N.~Spagnolo, F.~Sciarrino, Photonic quantum information processing: a review. \emph{Reports on Progress in Physics} \textbf{82}~(1), 016001 (2018), \doi{10.1088/1361-6633/aad5b2}, \url{https://dx.doi.org/10.1088/1361-6633/aad5b2}.

\bibitem{Slussarenko_2019}
S.~Slussarenko, G.~J. Pryde, Photonic quantum information processing: A concise review. \emph{Applied Physics Reviews} \textbf{6}~(4), 041303 (2019), \doi{10.1063/1.5115814}, \url{https://doi.org/10.1063/1.5115814}.

\bibitem{Friis_2019}
N.~Friis, G.~Vitagliano, M.~Malik, M.~Huber, Entanglement certification from theory to experiment. \emph{Nature Reviews Physics} \textbf{1}~(1), 72--87 (2019), \doi{10.1038/s42254-018-0003-5}, \url{https://doi.org/10.1038/s42254-018-0003-5}.

\bibitem{Erhard_2020}
M.~Erhard, M.~Krenn, A.~Zeilinger, Advances in high-dimensional quantum entanglement. \emph{Nature Reviews Physics} \textbf{2}~(7), 365--381 (2020), \doi{10.1038/s42254-020-0193-5}, \url{https://doi.org/10.1038/s42254-020-0193-5}.

\bibitem{Pirandola_2020}
S.~Pirandola, \emph{et~al.}, Advances in quantum cryptography. \emph{Advances in Optics and Photonics} \textbf{12}~(4), 1012 (2020), \doi{10.1364/aop.361502}, \url{http://dx.doi.org/10.1364/AOP.361502}.

\bibitem{Xu_2020}
F.~Xu, X.~Ma, Q.~Zhang, H.-K. Lo, J.-W. Pan, Secure quantum key distribution with realistic devices. \emph{Rev. Mod. Phys.} \textbf{92}, 025002 (2020), \doi{10.1103/RevModPhys.92.025002}, \url{https://link.aps.org/doi/10.1103/RevModPhys.92.025002}.

\bibitem{Mattle_1996}
K.~Mattle, H.~Weinfurter, P.~G. Kwiat, A.~Zeilinger, Dense Coding in Experimental Quantum Communication. \emph{Phys. Rev. Lett.} \textbf{76}, 4656--4659 (1996), \doi{10.1103/PhysRevLett.76.4656}, \url{https://link.aps.org/doi/10.1103/PhysRevLett.76.4656}.

\bibitem{Harrow_2004}
A.~Harrow, P.~Hayden, D.~Leung, Superdense Coding of Quantum States. \emph{Phys. Rev. Lett.} \textbf{92}, 187901 (2004), \doi{10.1103/PhysRevLett.92.187901}, \url{https://link.aps.org/doi/10.1103/PhysRevLett.92.187901}.

\bibitem{Barreiro_2008}
J.~T. Barreiro, T.-C. Wei, P.~G. Kwiat, Beating the channel capacity limit for linear photonic superdense coding. \emph{Nature Physics} \textbf{4}~(4), 282--286 (2008), \doi{10.1038/nphys919}, \url{https://doi.org/10.1038/nphys919}.

\bibitem{Hu_2018}
X.-M. Hu, \emph{et~al.}, Beating the channel capacity limit for superdense coding with entangled ququarts. \emph{Science Advances} \textbf{4}~(7), eaat9304 (2018), \doi{10.1126/sciadv.aat9304}, \url{https://www.science.org/doi/abs/10.1126/sciadv.aat9304}.

\bibitem{Wang_2018}
J.~Wang, \emph{et~al.}, Multidimensional quantum entanglement with large-scale integrated optics. \emph{Science} \textbf{360}~(6386), 285--291 (2018), \doi{10.1126/science.aar7053}, \url{https://www.science.org/doi/abs/10.1126/science.aar7053}.

\bibitem{Alexeev_2021}
Y.~Alexeev, \emph{et~al.}, Quantum Computer Systems for Scientific Discovery. \emph{PRX Quantum} \textbf{2}, 017001 (2021), \doi{10.1103/PRXQuantum.2.017001}, \url{https://link.aps.org/doi/10.1103/PRXQuantum.2.017001}.

\bibitem{Arrazola_2021}
J.~M. Arrazola, \emph{et~al.}, Quantum circuits with many photons on a programmable nanophotonic chip. \emph{Nature} \textbf{591}~(7848), 54--60 (2021), \doi{10.1038/s41586-021-03202-1}, \url{https://doi.org/10.1038/s41586-021-03202-1}.

\bibitem{Chi_2022}
Y.~Chi, \emph{et~al.}, A Programmable Qudit-based Quantum Processor, in \emph{CLEO 2023} (Optica Publishing Group) (2023), p. SF1E.1, \doi{10.1364/CLEO_SI.2023.SF1E.1}, \url{https://opg.optica.org/abstract.cfm?URI=CLEO_SI-2023-SF1E.1}.

\bibitem{Chang_2025}
K.-C. Chang, \emph{et~al.}, Quantum teleportation of a silicon nanophotonic CNOT gate. \emph{Optica Quantum} \textbf{3}~(4), 381 (2025), \doi{10.1364/opticaq.554577}, \url{http://dx.doi.org/10.1364/OPTICAQ.554577}.

\bibitem{Magana_2019}
O.~S. Maga{\~n}a-Loaiza, R.~W. Boyd, Quantum imaging and information. \emph{Reports on Progress in Physics} \textbf{82}~(12), 124401 (2019), \doi{10.1088/1361-6633/ab5005}, \url{https://dx.doi.org/10.1088/1361-6633/ab5005}.

\bibitem{Moreau_2019}
P.-A. Moreau, E.~Toninelli, T.~Gregory, M.~J. Padgett, Imaging with quantum states of light. \emph{Nature Reviews Physics} \textbf{1}~(6), 367--380 (2019), \doi{10.1038/s42254-019-0056-0}, \url{https://doi.org/10.1038/s42254-019-0056-0}.

\bibitem{Kues_2017}
M.~Kues, \emph{et~al.}, On-chip generation of high-dimensional entangled quantum states and their coherent control. \emph{Nature} \textbf{546}~(7660), 622--626 (2017), \doi{10.1038/nature22986}, \url{https://doi.org/10.1038/nature22986}.

\bibitem{Joshi_2020}
C.~Joshi, \emph{et~al.}, Frequency-Domain Quantum Interference with Correlated Photons from an Integrated Microresonator. \emph{Phys. Rev. Lett.} \textbf{124}, 143601 (2020), \doi{10.1103/PhysRevLett.124.143601}, \url{https://link.aps.org/doi/10.1103/PhysRevLett.124.143601}.

\bibitem{Lu_2022}
H.-H. Lu, \emph{et~al.}, Bayesian tomography of high-dimensional on-chip biphoton frequency combs with randomized measurements. \emph{Nature Communications} \textbf{13}~(1), 4338 (2022), \doi{10.1038/s41467-022-31639-z}, \url{https://doi.org/10.1038/s41467-022-31639-z}.

\bibitem{Clementi_2023}
M.~Clementi, \emph{et~al.}, Programmable frequency-bin quantum states in a nano-engineered silicon device. \emph{Nature Communications} \textbf{14}~(1), 176 (2023), \doi{10.1038/s41467-022-35773-6}, \url{https://doi.org/10.1038/s41467-022-35773-6}.

\bibitem{Lu_2023}
H.-H. Lu, M.~Liscidini, A.~L. Gaeta, A.~M. Weiner, J.~M. Lukens, Frequency-bin photonic quantum information. \emph{Optica} \textbf{10}~(12), 1655--1671 (2023), \doi{10.1364/OPTICA.506096}, \url{https://opg.optica.org/optica/abstract.cfm?URI=optica-10-12-1655}.

\bibitem{Tittel_1998}
W.~Tittel, J.~Brendel, H.~Zbinden, N.~Gisin, Violation of Bell Inequalities by Photons More Than 10 km Apart. \emph{Phys. Rev. Lett.} \textbf{81}, 3563--3566 (1998), \doi{10.1103/PhysRevLett.81.3563}, \url{https://link.aps.org/doi/10.1103/PhysRevLett.81.3563}.

\bibitem{Marcikic_2004}
I.~Marcikic, \emph{et~al.}, Distribution of Time-Bin Entangled Qubits over 50 km of Optical Fiber. \emph{Phys. Rev. Lett.} \textbf{93}, 180502 (2004), \doi{10.1103/PhysRevLett.93.180502}, \url{https://link.aps.org/doi/10.1103/PhysRevLett.93.180502}.

\bibitem{Xie_2015}
Z.~Xie, \emph{et~al.}, Harnessing high-dimensional hyperentanglement through a biphoton frequency comb. \emph{Nature Photonics} \textbf{9}~(8), 536--542 (2015), \doi{10.1038/nphoton.2015.110}, \url{https://doi.org/10.1038/nphoton.2015.110}.

\bibitem{Martin_2017}
A.~Martin, \emph{et~al.}, Quantifying Photonic High-Dimensional Entanglement. \emph{Phys. Rev. Lett.} \textbf{118}, 110501 (2017), \doi{10.1103/PhysRevLett.118.110501}, \url{https://link.aps.org/doi/10.1103/PhysRevLett.118.110501}.

\bibitem{Jaramillo_2017}
J.~A. Jaramillo-Villegas, \emph{et~al.}, Persistent energy\&\#x2013;time entanglement covering multiple resonances of an on-chip biphoton frequency comb. \emph{Optica} \textbf{4}~(6), 655--658 (2017), \doi{10.1364/OPTICA.4.000655}, \url{https://opg.optica.org/optica/abstract.cfm?URI=optica-4-6-655}.

\bibitem{Imany_2019}
P.~Imany, \emph{et~al.}, High-dimensional optical quantum logic in large operational spaces. \emph{npj Quantum Information} \textbf{5}~(1) (2019), publisher Copyright: {\textcopyright} 2019, The Author(s)., \doi{10.1038/s41534-019-0173-8}.

\bibitem{Chang_2021}
K.-C. Chang, \emph{et~al.}, 648 Hilbert-space dimensionality in a biphoton frequency comb: entanglement of formation and Schmidt mode decomposition. \emph{npj Quantum Information} \textbf{7}~(1), 48 (2021), \doi{10.1038/s41534-021-00388-0}, \url{https://doi.org/10.1038/s41534-021-00388-0}.

\bibitem{Chang_2025_2}
K.-C. Chang, X.~Cheng, M.~C. Sarihan, C.~W. Wong, Recent advances in high-dimensional quantum frequency combs. \emph{Newton} \textbf{1}~(1), 100024 (2025), \doi{https://doi.org/10.1016/j.newton.2025.100024}, \url{https://www.sciencedirect.com/science/article/pii/S2950636025000167}.

\bibitem{Chang_2023}
K.-C. Chang, X.~Cheng, M.~C. Sarihan, C.~W. Wong, Towards optimum Franson interference recurrence in mode-locked singly-filtered biphoton frequency combs. \emph{Photon. Res.} \textbf{11}~(7), 1175--1184 (2023), \doi{10.1364/PRJ.483570}, \url{https://opg.optica.org/prj/abstract.cfm?URI=prj-11-7-1175}.

\bibitem{Chang_2024}
K.-C. Chang, X.~Cheng, M.~C. Sarihan, C.~W. Wong, Time-reversible and fully time-resolved ultra-narrowband biphoton frequency combs. \emph{APL Quantum} \textbf{1}~(1), 016106 (2024), \doi{10.1063/5.0180543}, \url{https://doi.org/10.1063/5.0180543}.

\bibitem{Brecht_2015}
B.~Brecht, D.~V. Reddy, C.~Silberhorn, M.~G. Raymer, Photon Temporal Modes: A Complete Framework for Quantum Information Science. \emph{Phys. Rev. X} \textbf{5}, 041017 (2015), \doi{10.1103/PhysRevX.5.041017}, \url{https://link.aps.org/doi/10.1103/PhysRevX.5.041017}.

\bibitem{Fabre_2020}
C.~Fabre, N.~Treps, Modes and states in quantum optics. \emph{Rev. Mod. Phys.} \textbf{92}, 035005 (2020), \doi{10.1103/RevModPhys.92.035005}, \url{https://link.aps.org/doi/10.1103/RevModPhys.92.035005}.

\bibitem{Serino_2023}
L.~Serino, \emph{et~al.}, Realization of a Multi-Output Quantum Pulse Gate for Decoding High-Dimensional Temporal Modes of Single-Photon States. \emph{PRX Quantum} \textbf{4}, 020306 (2023), \doi{10.1103/PRXQuantum.4.020306}, \url{https://link.aps.org/doi/10.1103/PRXQuantum.4.020306}.

\bibitem{Mair_2001}
A.~Mair, A.~Vaziri, G.~Weihs, A.~Zeilinger, Entanglement of the orbital angular momentum states of photons. \emph{Nature} \textbf{412}~(6844), 313--316 (2001), \doi{10.1038/35085529}, \url{https://doi.org/10.1038/35085529}.

\bibitem{Krenn_2014}
M.~Krenn, \emph{et~al.}, Generation and confirmation of a (100 × 100)-dimensional entangled quantum system. \emph{Proceedings of the National Academy of Sciences} \textbf{111}~(17), 6243--6247 (2014), \doi{10.1073/pnas.1402365111}, \url{https://www.pnas.org/doi/abs/10.1073/pnas.1402365111}.

\bibitem{Malik_2016}
M.~Malik, \emph{et~al.}, Multi-photon entanglement in high dimensions. \emph{Nature Photonics} \textbf{10}~(4), 248--252 (2016), \doi{10.1038/nphoton.2016.12}, \url{https://doi.org/10.1038/nphoton.2016.12}.

\bibitem{Bavaresco_2018}
J.~Bavaresco, \emph{et~al.}, Measurements in two bases are sufficient for certifying high-dimensional entanglement. \emph{Nature Physics} \textbf{14}~(10), 1032--1037 (2018), \doi{10.1038/s41567-018-0203-z}, \url{https://doi.org/10.1038/s41567-018-0203-z}.

\bibitem{He_2022}
C.~He, Y.~Shen, A.~Forbes, Towards higher-dimensional structured light. \emph{Light: Science \& Applications} \textbf{11}~(1), 205 (2022), \doi{10.1038/s41377-022-00897-3}, \url{https://doi.org/10.1038/s41377-022-00897-3}.

\bibitem{Li_2023}
Y.~Li, \emph{et~al.}, Two-Measurement Tomography of High-Dimensional Orbital Angular Momentum Entanglement. \emph{Phys. Rev. Lett.} \textbf{130}, 050805 (2023), \doi{10.1103/PhysRevLett.130.050805}, \url{https://link.aps.org/doi/10.1103/PhysRevLett.130.050805}.

\bibitem{Zia_2023}
D.~Zia, N.~Dehghan, A.~D'Errico, F.~Sciarrino, E.~Karimi, Interferometric imaging of amplitude and phase of spatial biphoton states. \emph{Nature Photonics} \textbf{17}~(11), 1009--1016 (2023), \doi{10.1038/s41566-023-01272-3}, \url{https://doi.org/10.1038/s41566-023-01272-3}.

\bibitem{Scarfe_2025}
L.~Scarfe, Y.~Zhang, E.~Karimi, Spatial-Mode Quantum Cryptography in a 545-Dimensional Hilbert Space (2025), \url{https://arxiv.org/abs/2503.22058}.

\bibitem{Hu_2020}
X.-M. Hu, \emph{et~al.}, Efficient Generation of High-Dimensional Entanglement through Multipath Down-Conversion. \emph{Phys. Rev. Lett.} \textbf{125}, 090503 (2020), \doi{10.1103/PhysRevLett.125.090503}, \url{https://link.aps.org/doi/10.1103/PhysRevLett.125.090503}.

\bibitem{Hu_2021}
X.-M. Hu, \emph{et~al.}, Pathways for Entanglement-Based Quantum Communication in the Face of High Noise. \emph{Phys. Rev. Lett.} \textbf{127}, 110505 (2021), \doi{10.1103/PhysRevLett.127.110505}, \url{https://link.aps.org/doi/10.1103/PhysRevLett.127.110505}.

\bibitem{Erker_2017}
P.~Erker, M.~Krenn, M.~Huber, Quantifying high dimensional entanglement with two mutually unbiased bases. \emph{{Quantum}} \textbf{1}, 22 (2017), \doi{10.22331/q-2017-07-28-22}, \url{https://doi.org/10.22331/q-2017-07-28-22}.

\bibitem{Schneeloch_2019}
J.~Schneeloch, C.~C. Tison, M.~L. Fanto, P.~M. Alsing, G.~A. Howland, Quantifying entanglement in a 68-billion-dimensional quantum state space. \emph{Nature Communications} \textbf{10}~(1), 2785 (2019), \doi{10.1038/s41467-019-10810-z}, \url{https://doi.org/10.1038/s41467-019-10810-z}.

\bibitem{Valencia_2020}
N.~Herrera~Valencia, \emph{et~al.}, High-{D}imensional {P}ixel {E}ntanglement: {E}fficient {G}eneration and {C}ertification. \emph{{Quantum}} \textbf{4}, 376 (2020), \doi{10.22331/q-2020-12-24-376}, \url{https://doi.org/10.22331/q-2020-12-24-376}.

\bibitem{Mahler_2013}
D.~H. Mahler, \emph{et~al.}, Adaptive Quantum State Tomography Improves Accuracy Quadratically. \emph{Phys. Rev. Lett.} \textbf{111}, 183601 (2013), \doi{10.1103/PhysRevLett.111.183601}, \url{https://link.aps.org/doi/10.1103/PhysRevLett.111.183601}.

\bibitem{Rambach_2021}
M.~Rambach, \emph{et~al.}, Robust and Efficient High-Dimensional Quantum State Tomography. \emph{Phys. Rev. Lett.} \textbf{126}, 100402 (2021), \doi{10.1103/PhysRevLett.126.100402}, \url{https://link.aps.org/doi/10.1103/PhysRevLett.126.100402}.

\bibitem{Gross_2010}
D.~Gross, Y.-K. Liu, S.~T. Flammia, S.~Becker, J.~Eisert, Quantum State Tomography via Compressed Sensing. \emph{Phys. Rev. Lett.} \textbf{105}, 150401 (2010), \doi{10.1103/PhysRevLett.105.150401}, \url{https://link.aps.org/doi/10.1103/PhysRevLett.105.150401}.

\bibitem{Brougham_2013}
T.~Brougham, S.~M. Barnett, Mutually unbiased measurements for high-dimensional time-bin--based photonic states. \emph{Europhysics Letters} \textbf{104}~(3), 30003 (2013), \doi{10.1209/0295-5075/104/30003}, \url{https://dx.doi.org/10.1209/0295-5075/104/30003}.

\bibitem{Torlai_2018}
G.~Torlai, \emph{et~al.}, Neural-network quantum state tomography. \emph{Nature Physics} \textbf{14}~(5), 447--450 (2018), \doi{10.1038/s41567-018-0048-5}, \url{https://doi.org/10.1038/s41567-018-0048-5}.

\bibitem{Sahoo_2020}
S.~N. Sahoo, S.~Chakraborti, A.~K. Pati, U.~Sinha, Quantum State Interferography. \emph{Phys. Rev. Lett.} \textbf{125}, 123601 (2020), \doi{10.1103/PhysRevLett.125.123601}, \url{https://link.aps.org/doi/10.1103/PhysRevLett.125.123601}.

\bibitem{Lee_2014}
C.~Lee, \emph{et~al.}, Entanglement-based quantum communication secured by nonlocal dispersion cancellation. \emph{Phys. Rev. A} \textbf{90}, 062331 (2014), \doi{10.1103/PhysRevA.90.062331}, \url{https://link.aps.org/doi/10.1103/PhysRevA.90.062331}.

\bibitem{Mirhosseini_2015}
M.~Mirhosseini, \emph{et~al.}, High-dimensional quantum cryptography with twisted light. \emph{New Journal of Physics} \textbf{17}~(3), 033033 (2015), \doi{10.1088/1367-2630/17/3/033033}, \url{https://dx.doi.org/10.1088/1367-2630/17/3/033033}.

\bibitem{Zhong_2015}
T.~Zhong, \emph{et~al.}, Photon-efficient quantum key distribution using time–energy entanglement with high-dimensional encoding. \emph{New Journal of Physics} \textbf{17}~(2), 022002 (2015), \doi{10.1088/1367-2630/17/2/022002}, \url{https://dx.doi.org/10.1088/1367-2630/17/2/022002}.

\bibitem{Liu_2024}
J.~Liu, \emph{et~al.}, High-dimensional quantum key distribution using energy-time entanglement over 242 km partially deployed fiber. \emph{Quantum Science and Technology} \textbf{9}~(1), 015003 (2023), \doi{10.1088/2058-9565/acfe37}, \url{https://dx.doi.org/10.1088/2058-9565/acfe37}.

\bibitem{Chang_2024_2}
K.-C. Chang, M.~C. Sarihan, X.~Cheng, Z.~Zhang, C.~W. Wong, Large-alphabet time-bin quantum key distribution and Einstein–Podolsky–Rosen steering via dispersive optics. \emph{Quantum Science and Technology} \textbf{9}~(1), 015018 (2023), \doi{10.1088/2058-9565/ad0f6f}, \url{https://dx.doi.org/10.1088/2058-9565/ad0f6f}.

\bibitem{Korzh_2020}
B.~Korzh, \emph{et~al.}, Demonstration of sub-3 ps temporal resolution with a superconducting nanowire single-photon detector. \emph{Nature Photonics} \textbf{14}~(4), 250–255 (2020), \doi{10.1038/s41566-020-0589-x}, \url{http://dx.doi.org/10.1038/s41566-020-0589-x}.

\bibitem{Colangelo_2023}
M.~Colangelo, \emph{et~al.}, Impedance-Matched Differential Superconducting Nanowire Detectors. \emph{Phys. Rev. Appl.} \textbf{19}, 044093 (2023), \doi{10.1103/PhysRevApplied.19.044093}, \url{https://link.aps.org/doi/10.1103/PhysRevApplied.19.044093}.

\bibitem{BertaChristandlColbeckRenesRenner2010}
M.~Berta, M.~Christandl, R.~Colbeck, J.~M. Renes, R.~Renner, The uncertainty principle in the presence of quantum memory. \emph{Nat. Phys.} \textbf{6}, 659{\textendash}662 (2010), \url{https://doi.org/10.1038/nphys1734}.

\bibitem{DevetakWinter2005}
I.~Devetak, A.~Winter, Distillation of secret key and entanglement from quantum states. \emph{Proc. R. Soc. Lond. A} \textbf{461}, 207{\textendash}235 (2005), \url{https://doi.org/10.1098/rspa.2004.1372}.

\bibitem{BertlmannFriis2023}
R.~A. Bertlmann, N.~Friis, \emph{Modern Quantum Theory {\textendash} From Quantum Mechanics to Entanglement and Quantum Information} (Oxford University Press, Oxford, U.K.) (2023), \url{https://doi.org/10.1093/oso/9780199683338.001.0001}.

\bibitem{BennettBrassardPopescuSchumacherSmolinWootters1996}
C.~H. Bennett, \emph{et~al.}, Purification of Noisy Entanglement and Faithful Teleportation via Noisy Channels. \emph{Phys. Rev. Lett.} \textbf{76}, 722--725 (1996), \doi{10.1103/PhysRevLett.76.722}, \url{https://link.aps.org/doi/10.1103/PhysRevLett.76.722}.

\bibitem{Steinlechner_2017}
F.~Steinlechner, \emph{et~al.}, Distribution of high-dimensional entanglement via an intra-city free-space link. \emph{Nature Communications} \textbf{8}~(1) (2017), \doi{10.1038/ncomms15971}, \url{http://dx.doi.org/10.1038/ncomms15971}.

\bibitem{Bennett_Brassard_1984}
C.~H. Bennett, G.~Brassard, {Quantum cryptography: Public key distribution and coin tossing}, in \emph{Proceedings of IEEE International Conference on Computers, Systems, and Signal Processing} (IEEE, India) (1984), p. 175.

\bibitem{Ekert_1991}
A.~K. Ekert, {Quantum cryptography based on Bell's theorem}. \emph{Phys. Rev. Lett.} \textbf{67}, 661--663 (1991), \doi{10.1103/PhysRevLett.67.661}, \url{https://link.aps.org/doi/10.1103/PhysRevLett.67.661}.

\bibitem{Usenko_2025}
V.~C. Usenko, \emph{et~al.}, Continuous-variable quantum communication (2025), \url{https://arxiv.org/abs/2501.12801}.

\bibitem{Kanitschar_2024}
F.~Kanitschar, A.~Bergmayr-Mann, M.~Pivoluska, M.~Huber, Harnessing high-dimensional temporal entanglement using limited interferometric setups. \emph{Physical Review Applied} \textbf{22}~(5) (2024), \doi{10.1103/physrevapplied.22.054054}, \url{http://dx.doi.org/10.1103/PhysRevApplied.22.054054}.

\bibitem{Scarani_2009}
V.~Scarani, \emph{et~al.}, The security of practical quantum key distribution. \emph{Rev. Mod. Phys.} \textbf{81}, 1301--1350 (2009), \doi{10.1103/RevModPhys.81.1301}, \url{https://link.aps.org/doi/10.1103/RevModPhys.81.1301}.

\bibitem{Tupkary_2025}
D.~Tupkary, S.~Nahar, P.~Sinha, N.~L{\"{u}}tkenhaus, Phase error rate estimation in {QKD} with imperfect detectors. \emph{{Quantum}} \textbf{9}, 1937 (2025), \doi{10.22331/q-2025-12-11-1937}, \url{https://doi.org/10.22331/q-2025-12-11-1937}.

\bibitem{CurrasLorenzo_2025}
G.~Currás-Lorenzo, M.~Pereira, S.~Nahar, D.~Tupkary, Security of quantum key distribution with source and detector imperfections through phase-error estimation (2025), \url{https://arxiv.org/abs/2507.03549}.

\bibitem{Kanitschar_2025b}
F.~Kanitschar, M.~Huber, Composable Finite-Size Security of High-Dimensional Quantum Key Distribution Protocols (2025), \url{https://arxiv.org/abs/2505.03874}.

\bibitem{Tupkary_2024}
D.~Tupkary, E.~Y.-Z. Tan, N.~L\"utkenhaus, Security proof for variable-length quantum key distribution. \emph{Phys. Rev. Res.} \textbf{6}, 023002 (2024), \doi{10.1103/PhysRevResearch.6.023002}, \url{https://link.aps.org/doi/10.1103/PhysRevResearch.6.023002}.

\bibitem{Kanitschar_2025}
F.~Kanitschar, M.~Huber, Practical Framework for Analyzing High-Dimensional Quantum Key Distribution Setups. \emph{Phys. Rev. Lett.} \textbf{135}, 010802 (2025), \doi{10.1103/PhysRevLett.135.010802}, \url{https://link.aps.org/doi/10.1103/PhysRevLett.135.010802}.

\bibitem{Doda_2021}
M.~Doda, \emph{et~al.}, Quantum Key Distribution Overcoming Extreme Noise: Simultaneous Subspace Coding Using High-Dimensional Entanglement. \emph{Physical Review Applied} \textbf{15}~(3) (2021), \doi{10.1103/physrevapplied.15.034003}, \url{http://dx.doi.org/10.1103/PhysRevApplied.15.034003}.

\bibitem{Bertlmann_2008}
R.~A. Bertlmann, P.~Krammer, Bloch vectors for qudits. \emph{Journal of Physics A: Mathematical and Theoretical} \textbf{41}~(23), 235303 (2008), \doi{10.1088/1751-8113/41/23/235303}, \url{http://dx.doi.org/10.1088/1751-8113/41/23/235303}.

\bibitem{Hoeffding_1963}
W.~Hoeffding, {Probability Inequalities for Sums of Bounded Random Variables}. \emph{{J. Am. Stat. Assoc.}} \textbf{58}~(301), 13--30 (1963), \url{http://www.jstor.org/stable/2282952}.

\bibitem{Kanitschar_2023}
F.~Kanitschar, I.~George, J.~Lin, T.~Upadhyaya, N.~Lütkenhaus, Finite-Size Security for Discrete-Modulated Continuous-Variable Quantum Key Distribution Protocols. \emph{PRX Quantum} \textbf{4}~(4) (2023), \doi{10.1103/prxquantum.4.040306}, \url{http://dx.doi.org/10.1103/PRXQuantum.4.040306}.

\bibitem{Christandl_2009}
M.~Christandl, R.~König, R.~Renner, {Postselection Technique for Quantum Channels with Applications to Quantum Cryptography}. \emph{Phys. Rev. Lett.} \textbf{102}~(2), 020504 (2009), \doi{10.1103/physrevlett.102.020504}.

\bibitem{Nahar_2024}
S.~Nahar, D.~Tupkary, Y.~Zhao, N.~L\"utkenhaus, E.~Y.-Z. Tan, Postselection Technique for Optical Quantum Key Distribution with Improved de Finetti Reductions. \emph{PRX Quantum} \textbf{5}, 040315 (2024), \doi{10.1103/PRXQuantum.5.040315}, \url{https://link.aps.org/doi/10.1103/PRXQuantum.5.040315}.

\end{thebibliography}
\bibliographystyle{sciencemag}

\newpage

\section*{Acknowledgments}
The authors acknowledge discussions with Charles Lim, Alexander Euk Jin Ling, and discussions on the superconducting nanowire single-photon detectors with Aaron Miller and Vikas Anant. 
\vspace{-5pt}
\paragraph*{Funding:}
This study is supported by the Army Research Office Multidisciplinary University Research Initiative (W911NF-21-2-0214), National Science Foundation under award numbers 1741707 (EFRI ACQUIRE), 1919355, 1936375 (QII-TAQS), and 2137984 (QuIC-TAQS). Part of this research is performed at the Jet Propulsion Laboratory, California Institute of Technology, under contract with NASA. N.K.H.L., F.K., P.E. and M.H. acknowledge support from the European Research Council (Consolidator grant “Cocoquest” 101043705), the European flagship on quantum technologies (“ASPECTS” consortium 101080167), from the European Commission (grant 'Hyperspace’ 101070168), and the Austrian Federal Ministry of Education, Science, and Research via the Austrian Research Promotion Agency (FFG) grant scheme Quantum Austria (projects 914033, 921415 and 914030). N.K.H.L. also acknowledges support from the Austrian Science Fund (FWF) [P36478]. F.K. gratefully acknowledges support from the Dieberger-Skalicky foundation. Views and opinions expressed are those of the authors only and do not necessarily reflect those of the European Union.
\vspace{-5pt}
\paragraph*{Author contributions:}
K.-C.C., P.E., and M.H. developed the idea. K.-C.C. designed the experiments. K.-C.C., and M.C.S. conducted the measurements. K.-C.C., M.C.S., N.K.H.L., F.K., K.E.A., P.E., and M.H contributed to the data analysis. K.-C.C., N.K.H.L., F.K., P.E., and M.H. contributed to theoretical calculations. D.I.L., J.H.K., Y.C., A.M., M.D.S., B.K., and M.S. contributed the low-jitter SNSPD detectors. M.C.S., K.E.A., A.A., P.E., M.H., and C.W.W. supported and discussed the studies. K.-C.C., N.K.H.L., F.K., P.E., M.H., and C.W.W. prepared the manuscript. All authors contributed to the discussion and/or revision of the manuscript.
\vspace{-5pt}
\paragraph*{Competing interests:}

The authors declare no competing interests.
\vspace{-5pt}
\paragraph*{Data and materials availability:}
The datasets generated and analyzed during this study are available from the corresponding authors upon reasonable request. Source data are provided with this paper.

%%%%%%%%%%%%%%%% END OF MAIN TEXT %%%%%%%%%%%%%%%

\newpage

%%%%%%%%%%%%%%%% START OF SUPPLEMENT %%%%%%%%%%%%%%%

% Figures, tables, equations and pages in the supplement are numbered S1, S2 etc.
\renewcommand{\thefigure}{S\arabic{figure}}
\renewcommand{\thetable}{S\arabic{table}}
\renewcommand{\theequation}{S\arabic{equation}}
\renewcommand{\thepage}{S\arabic{page}}
\setcounter{figure}{0}
\setcounter{table}{0}
\setcounter{equation}{0}
\setcounter{page}{1} % not 0 as \newpage already started a supplementary page
% References continue the numbering from the main text.

%%%%%%%%%%%%%%%% SUPPLEMENT TITLE PAGE %%%%%%%%%%%%%%%

\begin{center}
\section*{Supplementary Materials for\\ \scititle}

% Author list for the supplement
% Indicate the corresponding authors, but do NOT include institutions here
% It would be nice if the template auto-generated this, but doing so is complicated...
\author{
	% You can write out first names or use initials - either way is acceptable, but be consistent
	Kai-Chi Chang$^{\ast\dagger}$,
	Murat Can Sarihan$^{\dagger}$,
	Nicky Kai Hong Li$^{\ast\dagger}$, Florian Kanitschar$^{\dagger}$,\\
    Kemal Enes Akyuz$^{\dagger}$,
    Yujie Chen,
    Dong-Il Lee,
    Jin Ho Kang,
    Alwaleed Aldhafeeri,\\
    Andrew Mueller,
    Matthew D. Shaw,
    Boris Korzh,
    Maria Spiropulu,\\
    Paul Erker$^{\ast}$,
    Marcus Huber$^{\ast}$,
    Chee Wei Wong$^{\ast}$
    \and
	
	% Identify at least one corresponding author, with contact email address
	\small$^\ast$Corresponding author: uclakcchang@ucla.edu; kai.li@tuwien.ac.at;
    paul.erker@tuwien.ac.at; marcus.huber@tuwien.ac.at; cheewei.wong@ucla.edu\\
	% Joint contributions can be indicated like this
	\small$^\dagger$These authors contributed equally to this work.
    }

\end{center}

% Fill out the numbers for each type of supplementary material,
% and delete any lines that aren't applicable.
% These are just example numbers that don't match the rest of this template.
\subsubsection*{The Supplementary Materials include:}
Methods\\
Cross time-frequency basis measurements\\
Figures S1 and S2\\
Table S1\\
References \cite{Bertlmann_2008,Hoeffding_1963,Kanitschar_2023,Christandl_2009,Nahar_2024}

\newpage

%%%%%%%%%%%%%%%% MATERIALS AND METHODS %%%%%%%%%%%%%%%

\section*{Methods}
\subsection*{Experimental details} 
We employ a continuous-wave distributed Bragg reflector single-frequency laser (Thorlab DBR780PN) to pump a type-II phase-matched, single-spatial-mode periodically-poled potassium titanyl phosphate (PPKTP) waveguide (AdVR Inc.) at 1560 nm. A fiber polarization controller (FPC) positioned before the PPKTP waveguide optimizes the generation of orthogonally-polarized SPDC photons. Residual pump photons are removed using a long-pass filter (LPF) and an angle-mounted band-pass filter (BPF) with 95\% passband transmission (Semrock NIR01-1570/3). Finally, a polarizing beam splitter (PBS) separates the signal and idler photons, directing them to Alice and Bob. Then, we implement the random choice of measurements between temporal basis ($T_A-T_B$) and frequency-resolved basis measurements ($F_A-F_B$) with 50:50 fiber beam splitters. This symmetric configuration guarantees ample coincidence counts to establish time-frequency MUBs measurements. The $T_A-T_B$ bases correspond to direct detection of photon arrival-time from both parties, and the $F_A-F_B$ bases are the dispersive basis that is mutually unbiased with respect to the temporal states. For both measurements, we utilize arrival-time high-dimensional encoding. Alice and Bob independently measure the photon arrival-times. Both parties use $N$ consecutive time-bins to form a time frame. For frequency-resolved measurements, we use a pair of large dispersion modules, with $\pm$ 10,000 ps/nm DEM (DCM), and each of them has a loss of only  3 dB (Proximion). The effective frequency-resolution in our experiments can be obtained as FWHM timing jitter divided by the applied dispersion, which is  0.00329 nm (0.41 GHz), sizably smaller than the FWHM bandwidth of our SPDC source (250 GHz).\\

The coincidence counts from the $T_A-T_B$ bases are recorded by two low-jitter SNSPDs \cite{Colangelo_2023}. Recently impedance-matched differential SNSPDs have been developed to simultaneously achieve a practical active area for efficient coupling to a single-mode fiber and low-jitter operation. The two detectors used in this work featured optical stacks with a double anti-reflection coating above the nanowire, optimized for 1550 nm, resulting in approximately 80\% efficiency at this wavelength and a timing jitter of around 13.1 ps. Impedance-matching in SNSPDs significantly improves the signal-to-noise ratio of the readout, with system timing jitter around 15 ps. Using these low-jitter SNSPDs and our coincidence counting module (Picoharp 300), we observed a temporal cross-correlation peak with a FWHM of approximately 32.9 ps, as shown in Figure \ref{fig:Figure1}c inset. The broadening of the FWHM of the cross-correlation peak is due to the electronic jitter of our coincidence counting module. In the future, it is conceivable that we could enhance detector jitter by utilizing quicker superconducting materials and advancements in nanofabrication, potentially enabling the resolution of temporal correlations of SPDC photons at the fundamental limit of two-photon correlation time. On the other hand, for frequency-resolved measurements, we register coincidence counts from $F_A$ and $F_B$ bases via low-jitter SNSPDs. Here we observed a temporal cross-correlation peak with a FWHM of approximately 125.5 ps, as shown in Figure \ref{fig:Figure1}d inset. In this case, the broadening of the FWHM of the cross-correlation peak is due to the electronic jitter of our coincidence counting module, and the imperfect non-local dispersion cancellation of our large dispersion modules.

In Table \ref{tab:loss_contribution}, we provide comprehensive characterization of the optical loss for the whole measurement setup in main text Fig.\,1b. The dominant factors limiting the performance of high-dimensional entanglement certification and high-dimensional QKD are the detection jitter, dispersion imperfections, and optical loss of the system.

\begin{table}[h!]
    \centering
    \tabcolsep=5pt
    \begin{tabular}{|c||c|}
    % \hline
    \hline
    Source of loss & Loss in dB \\
        \hhline{|=||=|}
        SPDC output fiber coupling & 3 \\
        \hline
        Fiber bench & 0.97\\
        \hline
        Longpass Filter (LPF) & 0.8\\
        \hline
        Bandpass Filter (BPF) & 0.2\\
        \hline
        Fiber Polarization Controllers (FPCs) & 3 (1 each)\\
         \hline
         PBS & 1.1 \\
          \hline
         Fiber connector loss & 1 \\
          \hline
          Fiber BS & 3 each \\
          \hline
          DCM & 3.67 \\
          \hline
          DEM & 2.61 \\
          \hline
          Low-jitter detectors & 2 (1 each) \\
          \hline
    \end{tabular}
    \caption{Characterized sources of optical loss in the experimental setup.}
    \label{tab:loss_contribution}
\end{table}

\subsection*{High-dimensional entanglement witness}\label{app:EntWitness}
In this section, we provide more details of the certification techniques we use to observe high-dimensional entanglement \cite{Bavaresco_2018}. To begin, we examine the certification process for high-dimensional entanglement in a bipartite quantum system, where the total Hilbert space $\mathcal{H}_{AB} = \mathcal{H}_A \otimes \mathcal{H}_B$ of an a priori unknown quantum state $\rho$ has local dimensions $\text{dim}\mathcal{H}_A = \text{dim}\mathcal{H}_B= d$. To certify the Schmidt number (or the entanglement dimension) of $\rho$, we consider the fidelity $F(\rho,\Phi)$ with respect to the target quantum state $\vert \Phi \rangle$, which takes the form:\\[-6pt]
\begin{equation}
 F(\rho,\Phi) = \text{Tr} \left( \vert \Phi \rangle \langle \Phi\vert \rho \right) =\sum_{m,n=0}^{d-1} \lambda_{m} \lambda_{n} \langle mm \vert \rho\vert nn\rangle  
\end{equation}\\[-6pt]
with $\lambda_{n}$ being the corresponding Schmidt coefficients for the target quantum state $\vert \Phi \rangle$. The entanglement dimension can be lower bounded by considering the following inequality, which holds for any quantum state $\rho$ with Schmidt number at most $k\leq d$:\\[-6pt]
\begin{equation}
F(\rho,\Phi) \leq B_k(\Phi) \coloneqq \sum_{m=0}^{k-1} \lambda_m^2 \label{eq:SNtiltedBasisWitness}
\end{equation}\\[-6pt]
with $m\in\{0, \dots , d-1\}$ such that $\lambda_{m} \geq \lambda_{m'} \;\forall \; m < m'$. Hence, any quantum state with $F(\rho,\Phi) >B_k(\Phi)$ is incompatible with a Schmidt number of $k$ or less, thereby certifying a minimum entanglement dimension of $k+1$.\\[-5pt]

Therefore, the subsequent step involves experimentally determining the fidelity of the quantum state $F(\rho,\Phi)$. We utilize the respective matrix elements to lower bound the fidelity of the target state $F(\rho,\Phi)$ by first separating it into two parts, $F(\rho,\Phi) = F_1(\rho,\Phi) + F_2(\rho,\Phi)$, with \\[-6pt]
\begin{align}
F_1 (\rho,\Phi) &\coloneqq \frac{1}{d} \sum_m \langle mm \vert \rho \vert mm \rangle  \\
F_2(\rho,\Phi) &\coloneqq \frac{1}{d} \sum_{m \neq n} \langle mm \vert \rho \vert nn \rangle 
\end{align}\\[-6pt]
if the target state is the maximally entangled state $\vert \Phi \rangle = \tfrac{1}{\sqrt{d}}\sum_{m=0}^{d-1} \vert mm \rangle$. The part $F_1 (\rho,\Phi)$ can be directly extracted from measurements in one basis, while the part $F_2 (\rho,\Phi)$ can be lower bounded by $\tilde{F_2}(\rho,\Phi)$ using measurements in an additional basis, where 
\begin{align}
\tilde{F_2} \coloneqq &\sum_j \langle \tilde{j}\tilde{j}^* \vert \rho \vert \tilde{j}\tilde{j}^* \rangle - \frac{1}{d} \sum_{m,n} \langle mn \vert \rho \vert mn \rangle \nonumber\\
&- \sum_{m\neq m', m\neq, n \neq n', n'\neq m'} \tilde{\gamma}_{mm'nn'} \sqrt{\langle m'n' \vert \rho \vert m'n' \rangle \langle mn \vert \rho \vert mn \rangle }
\end{align}      
where the $\tilde{\gamma}_{mm'nn'}$ is given by: 
\begin{equation}
\tilde{\gamma}_{mm'nn'} = \begin{cases}
    0, \quad &\text{if}\; (m-m'-n+n')\text{mod}_d \neq 0, \\
    \tfrac{1}{d}, &\text{otherwise.}
\end{cases}  
\end{equation} 
Therefore, by measuring in two different bases, we can constrain the fidelity term $F_2 (\rho,\Phi)$, this, in turn, provides a lower bound $\tilde{F} (\rho,\Phi)$ for the fidelity $F (\rho,\Phi)$ in the Schmidt-number witness inequality presented in Eq.\;\eqref{eq:SNtiltedBasisWitness}, resulting in the following relationship that relates the fidelity lower bound to the upper bound for states with Schmidt number $k$:
\begin{equation}
\tilde{F}(\rho,\Phi) \leq F(\rho,\Phi) \leq B_k(\Phi).
\end{equation}
By employing this inequality as a witness, the entanglement dimension $d_{ent}$ that is certifiable is the maximal $k$ such that $\tilde{F}(\rho,\Phi) \geq B_k (\Phi)$.\\

\subsection*{Entanglement of distillation}\label{app:DistillEnt}

Next, we describe how we can lower bound the \textit{distillable entanglement} or \textit{entanglement of distillation} $E_D$ in our quantum systems using two measurement bases. First, let us recall definition of the conditional Shannon entropy:
\begin{equation}
H(A_i|B_i) = H\left( \{ p_{jk}^{(i)}\} \right) - H\left( \{ p_{j}^{(i)}\} \right)
\end{equation} 
where $ p_{jk}^{(i)}= \langle j^{(i)} k^{(i)} \vert \rho\vert j^{(i)} k^{(i)} \rangle_{AB}$ and $ p_{j}^{(i)}= \sum_k \langle j^{(i)} k^{(i)} \vert \rho\vert j^{(i)} k^{(i)} \rangle_{AB}$ with $i$ being the basis label. Knowing that these terms are tied to coincidence counts measured in any two bases, we can bound the distillable entanglement $E_D$ from below with \cite{BertaChristandlColbeckRenesRenner2010,DevetakWinter2005,BertlmannFriis2023}: 
\begin{equation}
    E_D \geq -\log_2\left( \max_{i,j}  \vert \langle i \vert \tilde{j} \rangle \vert^2  \right) - H(A_1|B_1) - H(A_2|B_2),\label{eq:EoD}
\end{equation}                  
where $ \max_{i,j}  \vert \langle i \vert j \rangle \vert^2 $ is the maximal overlap of elements of the two bases used (which would be $1/d$ in case of ideal MUBs, as presented in the main text).

\subsection*{Comparison of the required number of local measurement settings versus different local dimensions $d$ for different techniques}

\begin{figure}[b!]
    \centering
    \includegraphics[trim=0 10pt 0 20pt,width=0.75\linewidth]{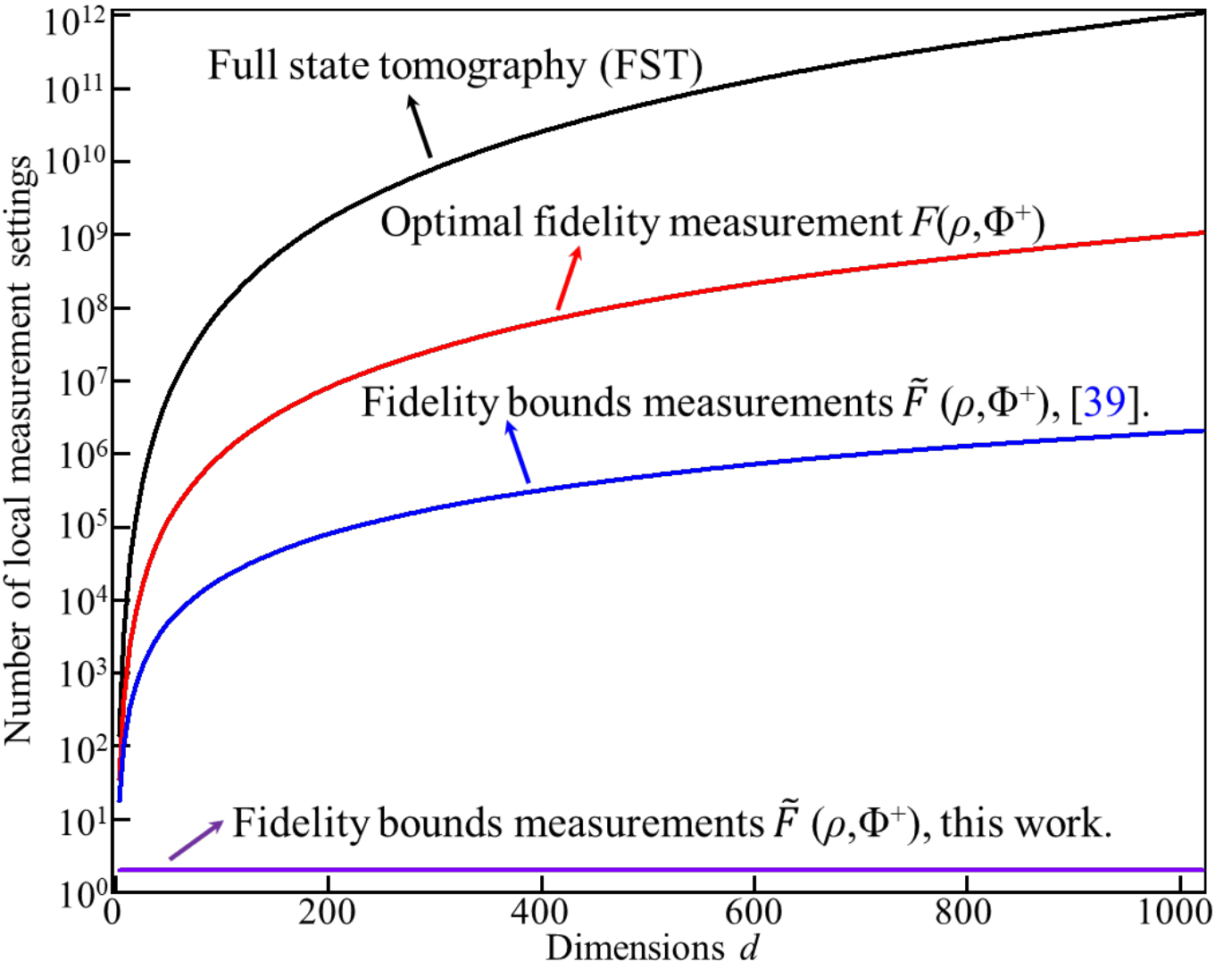}
    \caption{Comparison of the required number of local projective measurement settings in different local dimensions $d$ for different techniques. In this work, we highlight the constant number of measurement settings, since we only need a single setting for each of the $T_A-T_B$ and $F_A-F_B$ bases, independent of the dimensions $d$. Hence, our work represents many orders-of-magnitude improvement over traditional FST and prior literature \cite{Bavaresco_2018,Erker_2017,Valencia_2020,Bertlmann_2008}. We note that there are only a few fundamental limitations of our scheme: the number of measurable coincidence counts from the photon-pair source, loss in the time-to-frequency converter and other fiber components, as well as the timing jitter and detection efficiency of accessible single-photon detectors.}
    \label{fig:Data1}
\end{figure}
{\noindent}Extended Data Figure \ref{fig:Data1} compares the required number of local projective measurement settings used in this work with those required by other techniques across different dimensions $d$. For FST, $(d+1)^2 d^2$ local projective measurement settings are required \cite{Bavaresco_2018,Bertlmann_2008}, while optimal fidelity measurement $F(\rho,\Phi)$ requires $(d+1) d^2$ such measurements \cite{Bavaresco_2018}. More recently, it has been reported that only two measurement bases and $2d^2$ local projective measurement settings are sufficient to certify high-dimensional entanglement with Fidelity bounds $F(\rho,\Phi)$ \cite{Bavaresco_2018}. In this work, we highlight that only a constant number of measurement settings is required, since a single setting is sufficient for each of the $T_A-T_B$ and $F_A-F_B$ bases, independent of the dimensions $d$. Hence, our work represents many orders-of-magnitude improvement over traditional FST and prior literature \cite{Bavaresco_2018,Erker_2017,Valencia_2020,Bertlmann_2008}. For example, at $d=1021$, FST needs $\approx10^{12}$ local projective measurement settings, and prior works using the fidelity bound method need $\approx10^{6}$ settings \cite{Bavaresco_2018}, whereas our method requires only two measurement settings to certify high-dimensional entanglement of the quantum photonic state. We note that there are only a few fundamental limitations of our scheme: the number of measurable coincidence counts from the photon-pair source, loss in the time-to-frequency converter and other fiber components, as well as the timing jitter and detection efficiency of accessible single-photon detectors. \\

\subsection*{Composable security analysis}\label{app:QKDproof}
We adapt the adaptive-length \cite{Tupkary_2024} security argument from Ref. \cite{Kanitschar_2025b}. However, instead of using the witness-completion-approach from Ref. \cite{Kanitschar_2025}, we directly exploit the mutually unbiased bases measurement results and observe the correlation between Alice's and Bob's test rounds in the frequency basis $W = \sum_{i=0}^{d-1} P(ii| FF)$, where $P(ii|FF)$ is the probability that Alice and Bob obtain equal outcomes when both measure in the frequency basis \cite{Doda_2021}. Based on this observation, we find a statistical estimator $b_{\mathrm{stat}}(W)$, which is a high-probability lower bound to the Rényi entropy $H_{\alpha}(Z^n|C^nE^n)_{\rho}$ of the underlying quantum state, 
\begin{equation}
    \mathrm{Pr}\left[ b_{\mathrm{stat}}\left(W\right) \leq H_{\alpha}\left(Z^n|C^nE^n\right)_{\rho} \right] \geq 1-\epsilon_{\mathrm{AT}}.
\end{equation}
Here, $Z$ is the key register, $C$ is the transcript of the classical communication, $E$ denotes Eve's side information, and $n$ is the number of key rounds. In order to find such an estimator, we need to construct a set $V(W)$ which contains the unknown quantum state $\rho$ with high probability, $\mathrm{Pr}\left[ \tau_{AB} \in  V\left(W\right)\right] \geq 1-\epsilon_{\mathrm{AT}}$. Let $o_W$ be the observed statistics for observable $\hat{W} = \sum_{i=0}^{d-1} A_F^i \otimes B_F^i$ with $\{A_F^i\}_{i}$ and $\{B_F^i\}$ being Alice's and Bob's frequency basis measurements, we obtain
\begin{align}
    V(W) = \left\{ \sigma \in \mathcal{D}(\mathcal{H}_A\otimes\mathcal{H}_B\mathcal{H}_E:~ |\Tr{\hat{W}\sigma} - o_W| \leq \mu \right\},
\end{align}
where we obtain $\mu$ from Hoeffding's inequality \cite{Hoeffding_1963, Kanitschar_2023}
\begin{equation}
    \mu = \sqrt{\frac{2 ||\hat{W}||_{\infty}^2 }{k_W}  \ln\left(\frac{2}{\epsilon_{\mathrm{AT}}}\right)}.
\end{equation}
Here, $\epsilon_{\mathrm{AT}}$ is the security parameter associated with the statistical estimation procedure, $k_W$ is the number of rounds used to test $\hat{W}$. Then, the statistical estimator reads
\begin{equation}\label{eq:bStat}
        \begin{aligned}
            b_{\mathrm{stat}}\left(W\right):=& n \min_{\tau_{AB} \in V\left(W\right)} H_{\mathrm{min}}\left(X|E\right)_{\Phi_{\mathrm{var}}\left(\tau_{ABE} \right)}\\
            &- n (\alpha-1) \log_2^2\left( \dim(X)+1 \right)
        \end{aligned}
\end{equation}
where $1 < \alpha < 1+ \frac{1}{\log_2\left( 2\dim(X)+1 \right)}$.\\

Additionally, based on the observation and the communication transcript, we determine the error-correction leakage $\lambda^{\mathrm{EC}}(W, C)$. Then, the protocol conditioned on observing $W$ during the statistical testing procedure and conditioned on passing error-verification, hashes to a key length of
\begin{equation}\label{eq:VarLengthCollKRFormula}
    \ell\left(W\right) := \max\left\{0, b_{\mathrm{stat}}\left(W\right) - \lambda^{\mathrm{EC}}\left(W\right) - \theta(\alpha, \epsilon_{\mathrm{PA}}, \epsilon_{\mathrm{EV}}) \right\}
\end{equation}
where $\theta(\alpha,\epsilon_{\mathrm{PA}}, \epsilon_{\mathrm{EV}}) := \frac{\alpha}{\alpha-1} \left( \log_2\left( \frac{1}{4\epsilon_{\mathrm{PA}}} + \frac{2}{\alpha}\right) \right) + \left\lceil \log_2\left(\frac{1}{\epsilon_{\mathrm{EV}}} \right) \right\rceil$, using $\lambda^{\mathrm{EC}}\left(W, C\right)$ bits for error-correction is $\epsilon_{\mathrm{EV}}$-correct and is $\epsilon_{\mathrm{AT}} + \epsilon_{\mathrm{PA}}$-secure against i.i.d. collective attacks. \\

Using the postselection technique\cite{Christandl_2009, Nahar_2024}, we can lift security to general attacks. Once we proved security against collective attacks with security parameter $\epsilon_{\mathrm{PE}} + \epsilon_{\mathrm{AT}}$ and correctness parameter $\epsilon_{\mathrm{EV}}$ conditioned on obtaining $\vec{F}^{\mathrm{obs}}$ during acceptance testing and passing the error-verification, the protocol is secure against coherent attacks with security parameter $g_{n,x} \left( \sqrt{8 (\epsilon_{\mathrm{PE}} + \epsilon_{\mathrm{AT}})} + \frac{\tilde{\epsilon}}{2} \right)$, if the key is hashed to a length of
\begin{equation}\label{eq:VarLengthCohKRFormula}
\begin{aligned}
&\ell\left(\Vec{F}^{\mathrm{obs}}\right) :=\\
    & \max\left\{0, b_{\mathrm{stat}}\left(\Vec{F}^{\mathrm{obs}}\right) - \lambda^{\mathrm{EC}}\left(\Vec{F}^{\mathrm{obs}}\right) - \theta(\alpha, \epsilon_{\mathrm{PA}}, \epsilon_{\mathrm{EV}})  - 2 \log_2\left(g_{n,x}\right) - 2\log_2\left( \frac{1}{\tilde{\epsilon}}\right)\right\},
  \end{aligned}
\end{equation}
where $g_{n,x} = {n+x-1 \choose n}$ for $x = d_A^2 d_B^2$ and $\tilde{\epsilon}>0$ can be chosen freely. \\

Thus, it remains to determine the statistical estimator, i.e., to solve $\min_{\tau_{AB} \in V\left(W\right)}H_{\mathrm{min}}\left(X|E\right)_{\Phi_{\mathrm{var}}\left(\tau_{ABE} \right)}$. The present setup performs two MUB measurements, hence we may replace the semi-analytic duals method from Ref. \cite{Kanitschar_2025}, designed for evaluating $H_{\mathrm{min}}$ arbitrary setups, by a generalised version of the technique introduced in Ref. \cite{Doda_2021}, which exploits the high symmetry of mutually unbiased basis measurements. Consequently, we obtain for the statistical estimator
\begin{equation}
H_{\mathrm{min}}(Z|E)_{\rho} = \log_2d - 2\log_2\left(\sqrt{W - \mu}+\sqrt{(d-1)(1-W+\mu)}\right).
\end{equation}

We applied our security argument to the observed data. Therefore, we chose $\epsilon_{\mathrm{EC}} =\epsilon_{\mathrm{PA}}  = \epsilon_{\mathrm{AT}}  = \frac{1}{2} \times 10^{-10}$ and $\alpha = 1+ \frac{1}{\sqrt{n}}$, leading to a total security parameter of  $\epsilon_{\mathrm{sec}}= 10^{-10}$.

\section*{Cross time-frequency basis measurements}
\label{app:supp1}

\begin{figure}[h!]
\centering
\includegraphics[width=\textwidth]{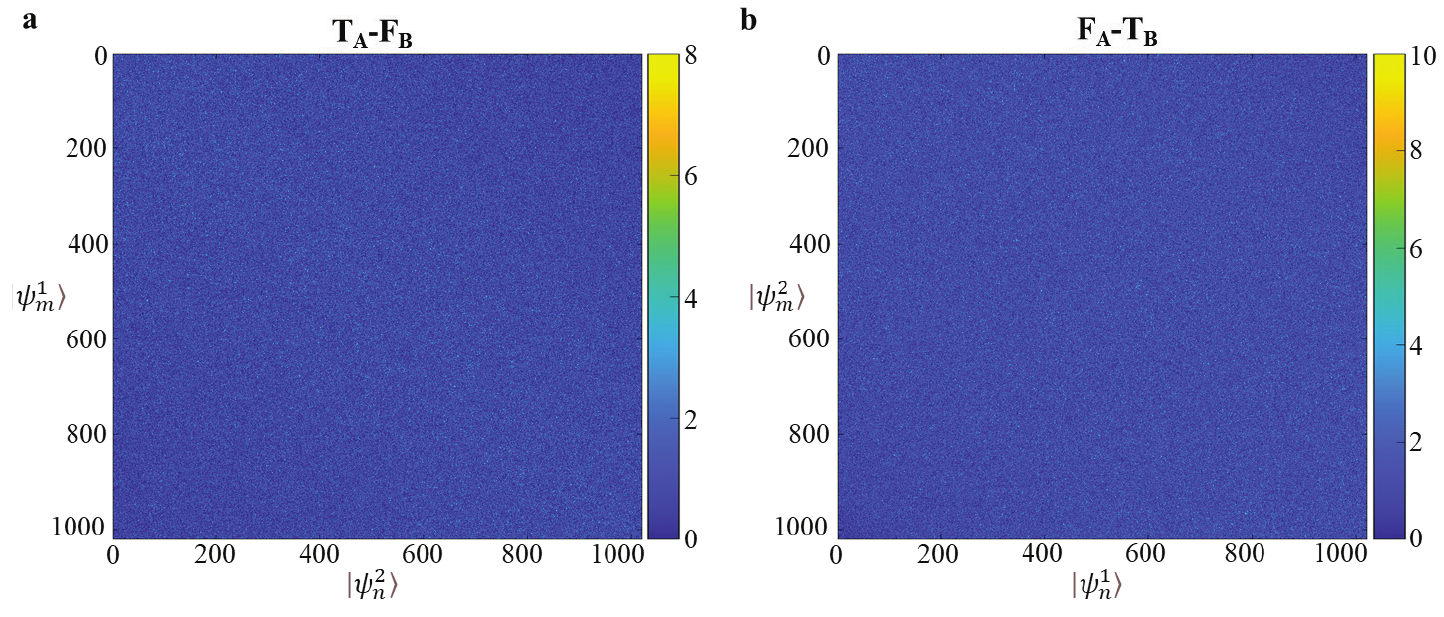}
% \end{center}
\caption{Time-frequency cross-basis measurements with dimension of 1021. |  a and  b, A 31-dimensional discretized JTI from $T_A$ and $F_B$ and $F_A$ and $T_B$ basis, respectively. The bin width $\tau$ and number of bins are $N$ is 10 ps and 1021, respectively. We can observe the near-uniform JTI with low coincidence counts. Here the duration of measured coincidence counting is 3 seconds, and no subtraction of background or accidental counts is performed.}
\label{fig:S1}
\end{figure}

Here we provide the experimental results and theoretical analysis of $T_A$ and $F_B$ and $F_A$ and $T_B$ basis. The A and B refer to Alice and Bob, respectively. For the cross-basis measurements, we perform them with experimental setup in the main text Figure \ref{fig:Figure1} b. Figures \ref{fig:S1} (a) and (b) are the measured 1021-dimensional discretized joint temporal intensity (JTI) from $T_A$ and $F_B$ and $F_A$ and $T_B$ basis, respectively. Here the bin width $\tau$ and number of bins is $N$ is 10 ps and 1021, respectively. We can observe the near-uniform JTI with low coincidence counts. For Figure \ref{fig:S1}, the duration of measured coincidence counting is 3 seconds, and no subtraction of background or accidental counts is performed. \\

We also quantify how close are the time-frequency bases to be mutually unbiased by evaluating $\Delta M$, the \textit{normalized} Frobenius norm $\frac{1}{2}||\cdot||_F$ of the difference between the normalized time-frequency bases' correlation matrix $C_{\text{TF}}$ and the ideal correlation matrix for MUBs $C_{\text{MUBs}} = \frac{1}{d^2}\mathbf{1}_{d\times d}$ with $(\mathbf{1}_{d\times d})_{ij}=1$ for all $i,j\in\{1,\ldots,d\}$:
\begin{align}
    \Delta M\coloneqq \frac{1}{2}||C_{\text{TF}} - C_{\text{MUBs}}||_F = \frac{1}{2}\sqrt{\sum_{i,j=1}^d \left|(C_{\text{TF}})_{ij} - \frac{1}{d^2}\right|^2}.
\end{align}
The calculated $\Delta M$ for different local dimensions $d$, in which we certify entanglement and evaluate secure key rates, are shown in Table\;\ref{tab:FrobeniusNorms}.

\begin{table}[h!]
    \centering
    \tabcolsep=5pt
    \begin{tabular}{|c||c|c|c|c|c|c|c|c|c|c|c|}
    % \hline
    \hline
    $\bm{d}$ & $\bm{3}$ & $\bm{7}$ & $\bm{13}$ & $\bm{31}$ & $\bm{61}$ & $\bm{127}$ & $\bm{251}$ & $\bm{331}$ & $\bm{419}$ & $\bm{509}$ & $\bm{1021}$\\
        \hhline{|=||=|=|=|=|=|=|=|=|=|=|=|}
        $\bm{\Delta M}$ & 0.03 & 0.01 & 0.007 & 0.003 & 0.002 & 0.0008 & 0.0006 & 0.0007 & 0.0005 & 0.0005 & 0.0005\\
        \hline
         % \hline
    \end{tabular}
    \caption{The normalized Frobenius norm of the difference between the normalized time-frequency bases' correlation matrix $C_{\text{TF}}$ and the ideal correlation matrix for MUBs $C_{\text{MUBs}}$. This measure of the bases biasness is denoted by $\Delta M$, and is shown here (rounded to 1 significant figure) for various local dimensions $d$.}
    \label{tab:FrobeniusNorms}
\end{table}

\end{document}